\documentclass[11pt,preprint]{aastex}

\received{}
\shortauthors {Shaw et al.}
\shorttitle {Chemical Abundances in SMC Planetary Nebulae} 

\begin{document}

\title {A Detailed Look at Chemical Abundances in Magellanic Cloud Planetary Nebulae. I. The Small Magellanic Cloud}

\author {Richard A.~Shaw\altaffilmark{1}, 
Ting-Hui Lee\altaffilmark{1,2}, 
Letizia Stanghellini\altaffilmark{1}, 
James E. Davies\altaffilmark{1},
D.~An{\'i}bal ~Garc\'\i a-Hern\'andez\altaffilmark{3},
Pedro Garc\'\i a-Lario\altaffilmark{4},
Jos{\'e} V.~Perea-Calder\'on\altaffilmark{5},
Eva Villaver\altaffilmark{6},
Arturo Manchado\altaffilmark{3,7}, 
Stacy Palen\altaffilmark{8}, 
Bruce Balick\altaffilmark{9}
}
\altaffiltext{1}{National Optical Astronomy Observatory, 950 N.\ Cherry Avenue, Tucson, AZ 85719}
\altaffiltext{2}{Department of Physics \& Astronomy, Western Kentucky University, Bowling Green, KY 42101}
\altaffiltext{3}{Instituto de Astrof\'isica de Canarias, v\'{\i}a L\'actea s/n, La Laguna, E-38200 Tenerife, Spain; also CSIC, Spain}
\altaffiltext{4}{Herschel Science Centre, European Space Astronomy Centre, Research and Scientific Support Department of ESA, Villafranca del Castillo, P.~O.\ Box 50727 E-28080 Madrid, Spain}
\altaffiltext{5}{European Space Astronomy Centre, INSA S.~A., P.~O.\ Box 50727 E-28080 Madrid, Spain}
\altaffiltext{6}{Universidad Aut{\'o}noma de Madrid, Departamento de F{\'i}sica Te{\'o}rica C-XI, 28049 Madrid, Spain}
\altaffiltext{7}{9 Consejo Superior de Investigaciones Cientificas, E-28006 Madrid, Spain}
\altaffiltext{8}{Physics Department, Weber State University, Ogden, UT 84408}
\altaffiltext{9}{Astronomy Department, Univ. of Washington, Seattle, WA 98195}

\email{shaw@noao.edu, ting-hui.lee@wku.edu, letizia@noao.edu, jdavies@noao.edu, agarcia@iac.es, Pedro.Garcia-Lario@sciops.esa.int, Jose.Perea@sciops.esa.int, eva.villaver@uam.es, amt@iac.es, spalen@weber.edu, balick@astro.washington.edu}


\begin{abstract} 

We present an analysis of elemental abundances of He, N, O, Ne, S, and Ar in Magellanic Cloud planetary nebulae (PNe), and focus initially on 14 PNe in the Small Magellanic Cloud (SMC). We derived the abundances from a combination of deep, high dispersion optical spectra, as well as mid-infrared (IR) spectra from the \textit{Spitzer Space Telescope}. A detailed comparison with prior SMC PN studies shows that significant variations among authors of 
relative emission line flux determinations 
lead to systematic discrepancies in derived elemental abundances between studies that are $\gtrsim0.15$~dex, in spite of similar analysis methods.
We used ionic abundances derived from IR emission lines, including those from ionization stages not observable in the optical, to examine the accuracy of some commonly used recipes for ionization correction factors (ICFs). These ICFs, which were developed for ions observed in the optical and ultraviolet, relate ionic abundances to total elemental abundances. We find that most of these ICFs work very well even in the limit of substantially sub-Solar metallicities, except for PNe with very high ionization. 
Our abundance analysis shows enhancements of He and N that are predicted from prior dredge-up processes of the progenitors on the AGB, as well as the well known correlations among O, Ne, S, and Ar that are little affected by nucleosynthesis in this mass range. 
We identified MG~8 as an interesting limiting case of a PN central star with a $\approx3.5$~M$_\sun$ progenitor in which hot-bottom burning did not occur in its prior AGB evolution. 
We find no evidence for O depletion in the progenitor AGB stars via the O-N cycle, which is consistent with predictions for lower-mass stars. 
We also find low S/O ratios relative to SMC \ion{H}{2} regions, with a deficit comparable to what has been found for Galactic PNe. 
Finally, the elemental abundances of one object, SMP-SMC~11, are more typical of SMC \ion{H}{2} regions, which raises some doubt about its classification as a PN. 

\end{abstract}


\section {Introduction} 

Studies of planetary nebulae (PNe) in the Magellanic Clouds over the past two decades have lead to substantial progress in understanding the late stages of stellar evolution, particularly the nucleosynthesis and mixing processes in their progenitor AGB stars that lead to the enrichment of He, C, and N in the interstellar medium. 
PNe in the Magellanic Clouds (MCPNe) are well suited to this endeavor in that their average metallicity, being substantially lower than Solar, constrains and challenges stellar evolution theory in ways not easy to test in other environments. Moreover, the progenitors of MCPNe span a broad range of abundances, owing to both the large range in main-sequence masses (roughly 1--8~M$_\sun$, with a correspondingly large range of pre-AGB ages), and to the increase in the average abundances in the ISM of the host galaxy over time, mostly from the ejecta of supernovae. 
MCPNe are easy to identify because of their bright nebular emission lines, they are numerous, and on average their emission is little affected by interstellar extinction. With the aid of high-resolution imaging, MCPNe are usually resolved and both the nebula and its central star can be easily observed against an often complex background of stellar and diffuse nebular emission. This set of observables and a known distance is a combination that cannot be matched for any other sample with today's technology. 
A good deal of spectroscopic data have been published on MCPNe in the past two decades, mostly in the optical, by  
\citet{MBC88}, 
\citet{MD91a, MD91b}, 
\citet{Vass_etal92}, and \citet[][who also included UV spectroscopy]{LD_96}; these data were compiled and augmented by \citet{LD_06}. 
More recently \citet{Stang_etal02}, \citet{Stang_etal03}, and \citet{Shaw_etal06} published \textit{HST}/STIS slitless optical and UV spectra; additional optical spectroscopy was published by \citet{IMC07}; 
and \citet{Stang_etal09} published carbon abundances based on UV spectroscopy. In all, abundances in more than 180 MCPNe have been derived and analyzed. 

The most comprehensive analysis of abundance patterns among MCPNe to date is that of \citet[][hereafter, LD06]{LD_06}, who explored the distinction between PNe of Type~I vs.\ non-Type~I \citep{Peimbert78, TPP97}, the possibility of O destruction by nucleosynthesis in the core of the progenitor AGB star via the O-N cycle, the efficiency of the third dredge-up at low metallicity, and the constancy of Ne, Ar, and S through AGB evolution, among other relationships. 
More recently, \citet{MCI09} also examined abundance patterns in MCPNe, using many of the above-cited data in addition to new observations by \citet[][hereafter, IMC07]{IMC07}. 
The various reported trends in elemental abundances in both studies had a good deal of scatter, which complicated the comparisons with predictions from stellar evolution theory for elemental yields. Indeed, LD06 noted that the spectrophotometry taken from the literature for several PNe was of rather poor quality, either because critical faint emission lines were not detected, or because the discrepancies with their own data were very large. In the end LD06 excluded many of the emission line fluxes from \citet{MBC88} and \citet{Vass_etal92} as unreliable; for the remaining PNe they attempted to remove some common sources of discrepancy between the various published abundances by re-deriving the abundances using their own determination of the nebular electron temperature ($T_e$) and density ($N_e$), a consistent set of supporting atomic data, and using a uniform set of prescriptions for ionization correction factors (ICFs). LD06 estimated average uncertainties in their final elemental abundances in the range 0.05~dex for He to as high as 0.5~dex for S. 

High accuracy of elemental abundance determinations from PNe is crucial for validating stellar evolution models, particularly the chemical yields, to useful precision. Achieving high accuracy demands great care in the observations and their calibration, a robust derivation of the abundances, and a thorough accounting of the observational uncertainties. Yet the subject of observational uncertainties is seldom discussed in quantitative depth in the literature, although \citet{Stasinska04} provides a fairly complete qualitative description of many of the key issues. A comparison of abundances of elements in individual nebulae among the above-cited papers shows differences as large as 1.0~dex or more, with differences of $\sim0.15$~dex being very common. This level of disagreement among authors generally has been noted before \citep[for MCPNe specifically by, e.g., LD06, and][]{MCI09}, and in some cases it is recognized as a significant problem \citep[see, e.g.,][]{PB99} for comparing observations to chemical yield predictions from theory. 
Close examination of the differences reveals a variety of causes, including discrepant emission lines fluxes (because of calibration errors, limited signal-to-noise, blending with other, closely spaced emission features, etc.), differences in adopted extinction constants, different adopted physical parameters (particularly $T_e$), different adopted atomic data (particularly collision strengths), and the adoption of different ICF formulae. These systematic discrepancies can form a cascade, where errors at each stage of the abundance calculation can magnify errors in subsequent stages in the chain. 

We undertook a program of spectrophotometry of MCPNe with the aim of improving the accuracy of abundance determinations to support a more precise comparison of the chemical yields that are predicted by stellar evolution theory. We are able to improve upon prior results by leveraging highly accurate, space-based emission line fluxes of brighter emission lines, adding new mid-IR fine structure emission lines from some ions (which are insensitive to $T_e$,
including those of [\ion{O}{4}], [\ion{Ne}{2}], [\ion{Ne}{3}], [\ion{Ne}{5}], [\ion{S}{3}], and [\ion{S}{4}]), 
and by obtaining additional, high-resolution ground-based optical spectra. In the process we have characterized many systematic difficulties of prior studies, and we have endeavored to quantify our uncertainties with great care. We focus in this paper on 14 PNe in the Small Magellanic Cloud (SMC). 
We describe the observing programs, data reduction, and emission line measurements in \S2. In \S3 we describe the derivation of the nebular physical diagnostics, and of ionic and elemental abundances, and compare them with those in the literature in detail. In \S4 we probe the systematic discrepancies and possible sources of error in elemental abundances between authors, and analyze our adopted abundances in the context of stellar evolution in a low metallicity environment. We conclude in \S5 with a summary and a few thoughts on future studies of abundances in emission line nebulae. 

\section {Observations}

Our strategy for optical observations was to obtain spectra of our target PNe that were deeper, had higher resolution, and broader wavelength coverage than were available in the literature. There are several advantages to this approach: Higher spectral resolution allows us to resolve closely spaced lines such as the [\ion{O}{2}] pair at $\lambda3726.0, 3728.8$; 
[\ion{Ne}{3}] $\lambda3967.4$ and H$\epsilon$ $\lambda3970.1$; 
[\ion{Ar}{4}] $\lambda4711.3$, \ion{He}{1} $\lambda4713.1$ and \ion{Ne}{4} $\lambda4715.4$; 
\ion{He}{2} $\lambda4859.3$ and H$\beta$ $\lambda4861.3$; 
\ion{He}{2} $\lambda6560.0$ and H$\alpha$ $\lambda6562.8$; 
the [\ion{N}{1}] pair at $\lambda5197.9, 5200.3$; 
as well as to distinguish nebular \ion{He}{2} $\lambda4685.7$ emission from broader stellar emission (when present); and to distinguish nebular from atmospheric emission (by virtue of the radial velocity of the targets). 
Thus in principle it is possible to make use of additional emission lines for diagnostics and to determine more accurate ionic abundances from lines that are often partially blended. Deeper exposures offer the possibility of employing less commonly used diagnostics, such as the faint, temperature-sensitive auroral lines of [\ion{O}{1}] $\lambda5577.3$ and [\ion{N}{2}] $\lambda5754.6$, or the density-sensitive doublet of [\ion{Ar}{4}] $\lambda\lambda4711.3, 4740.2$. Deeper exposures also increase the chance of detecting higher ionization stages of important elements, e.g., [\ion{Ar}{4}] and [\ion{Ne}{4}], which decreases the need for (or magnitude of) ICFs in computing elemental abundances. In practice, though, the low abundances of the SMC PNe mean that the emission lines of most ionic species are intrinsically fainter relative to H$\beta$ than their Galactic counterparts, so that deeper exposures are needed even for standard abundance determinations. Finally, the larger wavelength coverage, particularly the IR, allows us to determine abundances of ionization stages not observable in the optical, or to determine ionic abundances with greater accuracy, since they are far less affected by interstellar extinction and have little dependence on electron temperature. 

\subsection {Optical Spectra} 

The observations were obtained with the ESO Multi-Mode Instrument (EMMI) \citep{DDD86} on the 3.6-m New Technology Telescope (NTT) on La Silla, Chile in 2000 and  2003.  We observed our targets with both long-slit and echelle spectroscopy to obtain the necessary spectral coverage and resolution with the minimum number of configurations. The long-slit spectra were obtained with the blue arm, a slit length of 19\farcs5 which has a spatial resolution of $\sim0\farcs36$ pixel$^{-1}$, and grating \#3 which has a spectral resolution of R~$\approx3000$. We selected a wavelength range of 3500--3900~\AA\ for the 2000 observations, but adjusted this to 3700--4100~\AA\ for the 2003 observations to include a small wavelength overlap between the blue and red spectra. 
The echelle spectra were obtained with the red arm using grating \#9 and the cross-dispersing grism \#3.
The 2003 observations were obtained with an updated red CCD array with 2 chips, each with 2 independent amplifiers. The space between the detectors produces a gap in the spectral coverage at 4900--5000~\AA. 
The dispersion is $\approx0.12$~\AA~pixel$^{-1}$ ($\approx0.21$~\AA~pixel$^{-1}$ for the spectra taken in 2000); grating \#9 gives a spectral resolution of $R\approx8400$ (7700), corresponding to $\approx0.6$~\AA\ near the center of the wavelength range. 
 
 The observing log is presented in Table~\ref{tab:obsLog}, 
The log includes the target name using the catalog designations of \citet{SMP78} and \citet{MG92}, the UT date of the observation(s), the exposure times for each arm of the instrument, the width of the entrance slit, and the wavelength range covered. Note that most bright MCPNe are smaller than 1\farcs0, so that the slit size could be adjusted for the seeing and still intercept most of the nebular emission with little loss of spectral resolution. Because the airmass for our exposures ranged from about 1.2 to 1.5, we aligned the entrance slit with the parallactic angle for the beginning of the initial pair of exposures for each target. Even with this precaution, the parallactic angle changed by as much as 20\degr\ during a visit; regrettably, this telescope/instrument combination does not offer atmospheric dispersion correction, nor does it automatically rotate the slit to follow the parallactic angle. 

The long-slit data were calibrated following standard techniques, using 
IRAF\footnote{IRAF is distributed by the National Optical Astronomy Observatory, which is operated by the Association of Universities for Research in Astronomy, Inc., under cooperative agreement with the National Science Foundation.} 
software \citep{massey97}.  The bias and flat-field frames were combined by averaging them, with artifact rejection.  The spectra of the targets were then combined with artifact rejection, bias subtracted, trimmed, and flat-field corrected.  The task {\tt doslit}, within the \textit{noao.imred.specred} package, was used for spectral extraction, and wavelength and flux calibration \citep{massey92}.  The extracting aperture was 3\farcs6, which was more than enough to ensure that the entire spatial extent of the nebula was included. 

The echelle spectra were calibrated following the steps outlined in \cite{willmarth94}. 
The target spectra, standard star spectra, flat fields, and wavelength calibration files were first bias subtracted using an average bias frame, then trimmed to remove the overscan as well as the bluest orders that were too faint to trace. 
The photometric uniformity of the target spectra, standard star spectra, and wavelength calibration files were then corrected with the normalized flat.
The extraction aperture for the targets was 3\farcs5; sky emission was extracted on either side of each order with aperture sizes of 3\farcs2 (for the 2003 observations) or 2\farcs2 (for the 2000 observations), which were separated from the target apertures with a gap of $\approx1$\arcsec.  Th--Ar comparison arcs provided an accurate  wavelength calibration. 
A correction for atmospheric extinction was made using average values for the observing site, and a response function was constructed for each order using standard star spectra obtained on each night. 
The flux-calibrated standard star spectra were then checked for order-to-order discontinuities, and the extraction and sensitivity function determination were refined as necessary to ensure an accurate calibration of the target spectra. 

Although the long-slit and echelle data for a given target were usually obtained consecutively, the blue arm (long-slit) spectra needed to be placed on the same intensity scale as the echelle (red-arm) spectra. This scaling is necessary because the spectra were taken in conditions that were not always photometric or with unstable seeing, so that grey extinction, slit losses, and other factors can in principle be different. Our original intent was to use the emission lines common to both spectra to establish the scale factor. Unfortunately these lines are weak and few in number, and more importantly the sensitivity in the bluest orders of the red arm is low and declining so steeply with decreasing wavelength that scale factors so derived were unreliable. Comparisons with published data for this purpose proved unreliable as well (see \S2.2). In the end we derived the blue/red scale factors using the H Balmer decrement, such that the H$\gamma$, H$\delta$, and H$\epsilon$ intensities were in good agreement with that predicted from the $I$(H$\beta$)/$I$(H$\alpha$) ratio, and the derived electron temperature and density. The adopted blue/red scale factors are given in the last column of Table~\ref{tab:obsLog}, and are usually within 10\% of unity. 

\subsection {Optical Emission Line Intensities} 

We measured the emission line intensities in all the NTT spectra using the IRAF \textit{splot} routine. We directly integrated the flux above the local continuum wherever possible, except for close, nearly blended lines where we cross-checked the individual line fluxes with Gaussian deblending. The physical size of the red-arm detector is sufficiently large that some emission lines appear on opposite ends of two adjacent echelle orders. In these cases we found that the fluxes measured in each order agreed to within 10\% in most cases; ultimately we selected the flux from the line closest to the center of the order, both because of the higher signal-to-noise ratio and because the order trace was better defined near the order centers. The level of agreement between orders sets a lower limit to the uncertainty of all the fluxes. 
The final nebular emission line intensities are presented in Table~\ref{tab:optFlux}, after application of the red/blue scale factors but uncorrected for interstellar reddening. In successive columns we give the identification of the ion responsible for the emission, the rest wavelength, and the relative line intensity for each target, normalized to $I$(H$\beta$)=100. Note that the intensity of [\ion{O}{3}] $\lambda4959$ is missing in most cases: this line falls in the gap between the detectors in the echellograms obtained in 2003. 

\begin {figure}
\epsscale{0.8}
\plotone {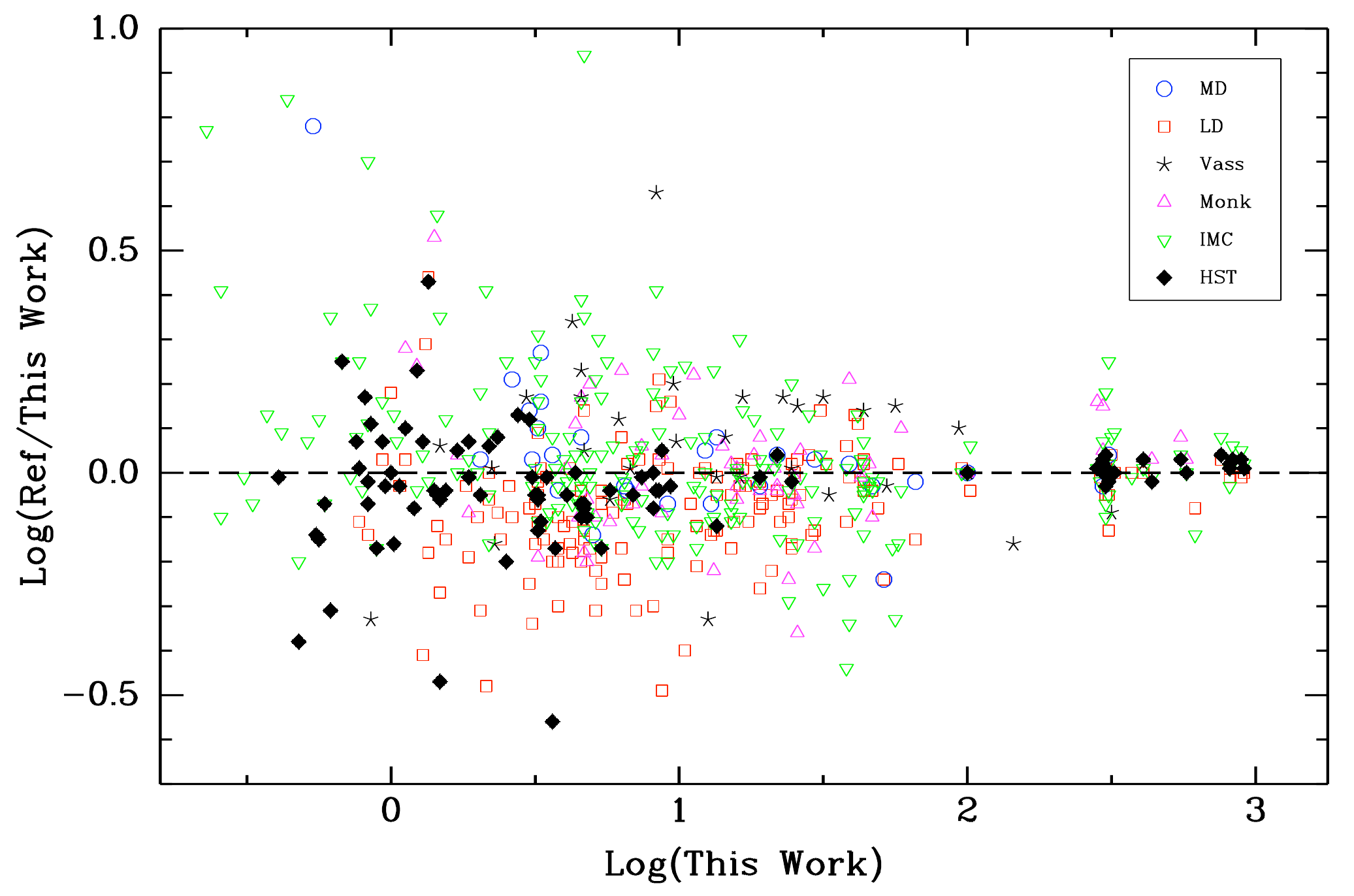}
\figcaption[f1.pdf]{ 
Comparison of published emission line intensities (Ref) in common with those reported in this paper, with the 1:1 relation indicated ({\it dashed line}). Symbol key: 
\citet{MD91a, MD91b} ({\it circles}); 
\citet{LD_96, LD_06} ({\it squares}); 
\citet{Vass_etal92} ({\it asterisks}); 
\citet{MBC88} ({\it triangles}); 
\citet{IMC07} ({\it inverted triangles}); 
\textit{HST} spectra from \citet{Stang_etal03} ({\it filled diamonds}). [\textit{See the electronic edition of the Journal for a color version of this figure.}]
\label{FlxCmp}}
\end {figure}

Every one of our targets has been observed spectroscopically by others, although usually with some combination of lower resolution, shallower exposures, or lesser wavelength coverage. We have compared our optical emission line intensities with those published by several other investigators: 
\citet{MBC88}, \citet{MD91a, MD91b}, \citet{Vass_etal92}, \citet{LD_96, LD_06}, \citet[][hereafter, IMC07]{IMC07}, 
as well as those from the \textit{HST}/STIS spectra of \citet{Stang_etal03}. The comparison data to which we attach the most weight are the \textit{HST}/STIS slitless spectra, owing to their superior absolute and relative calibration. But even these STIS data have limitations, particularly for targets larger than $\sim1\farcs2$ in size because of the spatial overlap between H$\alpha$ and nearby lines, and because fluxes from extended targets generally had lower signal-to-noise ratios. Therefore the STIS fluxes from MG~8 and SMC~11 have been excluded from the following comparison. 
Figure~\ref{FlxCmp} shows moderately good agreement between our fluxes and those in the literature, with the best agreement for strong lines and an increase in dispersion at fainter flux levels. A few trends are worth noting, however. First, the intensities from \citet{MD91a, MD91b} are systematically brighter than those presented here when the intensity we measure is $\lesssim5$\% of $I$(H$\beta$), with a dispersion that is quite large compared to their quoted errors. Second, the intensities from \citet{LD_96, LD_06} are 17\% fainter than ours when the intensity we measure is less than that of H$\beta$, although the agreement for brighter lines is excellent. Third, the intensities published by IMC07 show a large dispersion relative to ours, well beyond their quoted errors, and trend systematically brighter by $\sim$8\% when the intensity we measure is less than $\sim$5\% that of H$\beta$. It should be noted, however, that IMC07 adopted mean emission line intensities that were averaged over available data in the literature. This complicates the comparison because each emission line was determined from a different combination of observations and literature values. Thus, systematic errors, while undoubtedly present, are very difficult to characterize. 

The excellent agreement between the line intensities presented here and those from the \textit{HST}/STIS data (within 3\% in the mean) suggests that our intensities are the more reliable ground-based measurements. Based upon the level of internal consistency in our line intensity measurements and the above comparison of these intensities with values in the literature, we assign an uncertainty to our data of 10\% for intensities brighter than 10.0 (on the scale of $I$(H$\beta)=100$ prior to correction for interstellar reddening), 20\% for intensities between 2.0 and 10.0, and 40\% for intensities fainter than 2.0. Intensities fainter than 0.5 probably have uncertainties approaching 100\%. 

\subsection {Infrared Emission Line Intensities}

Infrared spectra of several of our SMC PNe were obtained with the Infrared Spectrograph \citep{Houck04} on-board the \textit{Spitzer Space Telescope} between 2005 July and November, as a part of General Observer program 20443. 
The spectra from this shallow survey are all low dispersion (R$\sim100$) covering the 5--38~$\mu$m range, and of relatively short duration (less than 2 min total exposure time per spectral region). \citet{Stang_etal07} describe the observing program and the data reduction techniques. The nebular emission line fluxes were measured with a Gaussian line-fitting routine in the SMART\footnote{SMART was developed by the IRS Team at Cornell University and is made available through the \textit{Spitzer} Science Center at Caltech.} package. Many expected emission lines were either below the detection limit, or were embedded within strong solid-state features. Where possible, we derived upper limits by computing the flux of a fictitious Gaussian emission line with a width typical of narrow emission features at the expected wavelength, and a height equal to three times the RMS noise in the local continuum. 
Our data were supplemented with emission line fluxes from \citet{BernardSalas08}, whose similar GTO program included \textit{Spitzer}/IRS spectra of two targets in our sample. 
The emission line intensities from all the \textit{Spitzer} spectra, which included all targets except for MG~8, MG~13, and SMP~9,
are presented in Table~\ref{tab:irFlux}. In successive columns we give the identification of the ion responsible for the emission, the rest wavelength in microns, the reddening function $f$($\lambda$), and the relative line intensity for each target, uncorrected for interstellar reddening and normalized to $I$(H$\beta$)=100. For subsequent analysis we correct the intensities for interstellar reddening in the standard way by multiplying them by $10^{cf(\lambda)}$, where $c$ is the logarithmic extinction at H$\beta$. 
The log of the absolute H$\beta$ fluxes, given in the first row, are from \citet{Stang_etal03}. Although there are no independent IR emission line flux measurements for comparison with these data, for the purpose of computing abundance uncertainties we adopt our statistical flux uncertainties, which are of order 10\% for most lines. 

\section {Chemical Abundances}

\subsection {Physical Diagnostics}

Determining the interstellar extinction and the physical diagnostics (gas density and temperature) is the first step in deriving the elemental abundances of the nebular gas. We have adopted the interstellar extinction constants from the accurate spectrophotometry of \citet{Stang_etal03}, with the exception of those for MG~13, for which the H$\alpha$ flux is not available, and SMP~11, where the angular extent of the nebula is large enough that the \textit{HST} fluxes at H$\alpha$ and [\ion{N}{2}] $\lambda\lambda$6548,~6583 suffered considerable overlap. In these cases we derived the extinction constant, $c$, from our measurements of the Balmer decrement in the NTT spectra, assuming an intrinsic $I$(H$\alpha$)/$I$(H$\beta$) ratio of 2.85 for case~B recombination. 
We used ratios of collisionally excited emission lines and the \textbf{nebular} software package \citep{SD95, Shaw_etal98} to determine the nebular electron densities ($N_e$) and temperatures ($T_e$). Specifically, we constructed traditional diagnostic diagrams for each object, plotting curves in the $N_e$, $T_e$ plane that were consistent with the observed (dereddened) ratios of 
[\ion{O}{2}] $I$($\lambda3726.0$)]/$I$($\lambda3728.8$), 
[\ion{Ar}{4}] $I$($\lambda4711.3$)]/$I$($\lambda4740.2$), and 
[\ion{S}{2}] $I$($\lambda6716.5$)]/$I$($\lambda6730.9$) for density; and 
[\ion{N}{2}] [$I$($\lambda6548.0$)+$I$($\lambda6583.4$)]/$I$($\lambda5754.6$) 
and [\ion{O}{3}] $I$($\lambda5006.8$)/$I$($\lambda4363.2$) 
for temperature. 
A few other diagnostic curves were plotted as well, including the optical temperature diagnostics 
[\ion{O}{2}] $I$($\lambda3727$)]/$I$($\lambda7325$) 
and [\ion{S}{2}] $I$($\lambda4072$)]/$I$($\lambda6725$), 
but in practice the uncertainties in these ratios are too large to be useful for this purpose. 
They were instead used as a check on the accuracy of the flux calibration and the extinction correction, for which the long wavelength baseline is a strong advantage. 
In the infrared, it is possible to determine $N_e$ from the IR ratios of 
[\ion{S}{3}] $I$($\lambda18.68~\mu$m)]/$I$($\lambda33.64~\mu$m), and 
[\ion{Ne}{5}] $I$($\lambda14.29~\mu$m)]/$I$($\lambda29.23~\mu$m), which are sensitive to densities of a few $\times 10^3$ to $\sim10^5$~cm$^{-3}$. In practice, the red-ward line fluxes are usually upper limits, so at best we can only use them here to exclude the possibility of moderate- to high-ionization regions with densities $\ga10^4$~cm$^{-3}$. Temperature indicators from ratios of optical to IR emission lines can also be used in principle: those available here are 
[\ion{S}{3}] $I$($\lambda6312.1$)]/$I$($\lambda18.68~\mu$m), and 
[\ion{Ne}{3}] $I$($\lambda3868.7$)]/$I$($\lambda15.56~\mu$m). In practice the [\ion{S}{3}] $\lambda6312.1$ line is too weak for a reliable temperature, although the median of the differences between the $T_e$ derived from this ion and that from [\ion{O}{3}] is less than 100~K. The [\ion{Ne}{3}] lines, while strong, suffer from a systematic discrepancy between the IR and the optical emission lines which is discussed in \S3.3.1. Therefore, the temperatures derived from IR lines are not sufficiently reliable for this purpose, but do serve as another check on the relative flux calibration and reddening correction. 

The derived physical diagnostics are presented in Table~\ref{tab:Diag}, where the first two columns give the target (common) designation and the adopted extinction constant. The next three columns give $N_e$ as derived from three diagnostic ratios, with the adopted density in the following column. Note that the adopted density is generally taken from the [\ion{O}{2}] ratio because the signal-to-noise ratio is always higher in these lines; in addition, the [\ion{S}{2}] ratios are in some cases near the high density limit, meaning that a small uncertainty in the flux ratio yields a very high uncertainty in the inferred density. The [\ion{Ar}{4}] lines are extremely weak and are therefore unreliable as a density diagnostic. 
The next two columns give $T_e$ as derived from the available diagnostics; formal uncertainties were also determined from the uncertainties in the flux measurements, which were taken from the prescription in \S2.2. In all but one case the error bar of $T_e$([\ion{N}{2}]) overlaps with $T_e$([\ion{O}{3}]), and in two cases $T_e$([\ion{N}{2}]) $>$ $T_e$([\ion{O}{3}]). In general $T_e$([\ion{N}{2}]) would be expected to be more representative of the low-ionization zone, which would contain most singly-ionized species. The weakness of these lines in many objects, which is largely a result of a low N abundance, means that they seldom yielded $T_e$ with a sufficiently small uncertainty to be useful. 
The adopted temperatures, given in the last two columns in the Table, are taken from [\ion{N}{2}] for the low-ionization region (when reliable), and from [\ion{O}{3}] for the medium-ionization region; we adopt $T_e$([\ion{O}{3}]) when $T_e$([\ion{N}{2}]) is not available or when it exceeds $T_e$([\ion{O}{3}]). 

\subsection {Ionic Abundances}

The abundance for He$^+$ was derived from the $\lambda5875.7$ recombination line. For these calculations we used the electron temperature as derived from [\ion{O}{3}], and the case~B recombination coefficients from \citet{Porter_etal05} which makes use of the latest radiative and collisional data. The abundance from the recombination lines at $\lambda4471.5$ and $\lambda6678.2$, was computed as well, but these lines are a factor of $\sim3$ weaker (and typically only a few percent of H$\beta$), with correspondingly greater uncertainty. The average of the abundance as derived from $\lambda4471.5$ and $\lambda6678.2$, where available, generally agreed well with that from $\lambda5875.7$. The He$^{+2}$ abundance was derived from the $\lambda4685.7$ recombination line, $T_e$([\ion{O}{3}]), and the recombination coefficients as tabulated in \citet{OF06}. The results are given in the first two rows of Table~\ref{tab:Ionic} for each nebula. 

The ionic abundances for collisionally excited lines were derived using the \textbf{nebular} software from the physical diagnostics determined above---where T$_e$ for the appropriate ionization zone was used \citep[see][]{SD95, Shaw_etal98}---and the emission line intensities from ions of N, O, Ne, S, and Ar as compiled in Tables~\ref{tab:optFlux} and \ref{tab:irFlux} and corrected for reddening using the \citet{Prevot_etal84} extinction curve for the SMC. We note that the \textbf{nebular} package has been updated with more recent energy levels and transition probabilities for various ions \citep[see the compilation by][for references]{Badnell_etal06}, including N$^{+2}$, O$^{+2}$, O$^{+3}$, Ne$^{+2}$, Ne$^{+4}$, and Ne$^{+5}$ which are of interest here. 
The ionic abundances relative to H$^+$ for collisionally excited ions are presented in Table~\ref{tab:Ionic} for each nebula. The abundances as computed from the mid-infrared lines are listed separately and annotated with ``IR'' in the first column, next to the ion identification. For some ions the abundance may be computed from either the optical or IR lines, which provides a useful comparison that will be discussed below.  The adopted ionization correction factor (see below) is also listed for each element following the abundances for the observed ions. 

\subsection {Elemental Abundances}

The elemental abundances of He, N, O, Ne, S, and Ar are presented in Table~\ref{tab:Abund}, based on the analysis presented in this section. 
The recombination lines of both ions of He are present in the optical spectra, so the elemental abundance is simply the sum of the abundances derived from each ion. He$^+$ is detected in all the PNe studied here, but often He$^{+2}$ is not. We note that the emission from \ion{He}{2} $\lambda4685.7$ in MG~8 is entirely of stellar origin: it is incorrectly attributed by \citet{Vass_etal92} as nebular emission. Further, \citet{Vass_etal92} give only an upper limit to the intensity of \ion{He}{1} $\lambda5875.7$ of $<5$, yet it is clearly present in our spectrum with an intensity of 14.4. Thus, the He abundance derived by LD06 based on the \citet{Vass_etal92} data is incorrect. 

\subsubsection {Ionization Correction Factors}

We derived total elemental abundances for N, O, Ne, S, and Ar, relative to H$^+$, by summing the abundances of the observed ions and correcting for unobserved ionization stages using ICFs. We generally adopted empirical relations for ICFs from \citet{KB94} for all elements except S, where we instead adopted the recipe from \citet{KH01}. However, some of the IR emission lines originate from ionization stages that are not observable in the optical, so that ICFs are sometimes not needed or, in the case of upper limits, can be confirmed. The adopted ICFs for each element are given in Table~\ref{tab:Ionic}, and the log elemental abundances are given in Table~\ref{tab:Abund} normalized to log~H$^+=12.0$. 

The first two ionization stages of oxygen (O$^+$ and O$^{+2}$) are detected in the optical spectra for all of the PNe in our sample. In the IR spectra O$^{+3}$ is detected in four objects, and tight upper limits are established for an additional seven objects. In these cases it is clear that the O abundance is properly obtained from the straight sum of the observed ionic abundances (i.e., ICF(O) $=1.0$), since higher ionization stages do not contribute significantly. For the three remaining objects where no IR spectrum is available, ICF(O) is computed from the He ionic abundances using Eq.~A9 from \citet{KB94}. 

For nitrogen, the only collisionally excited emission lines in the optical spectra are of N$^+$ (very weak recombination lines of \ion{N}{3} are detected in some objects), and no N lines are observable in our IR spectra. Although \citet{LD_96} published UV spectra from \textit{IUE}, the upper limits for N$^{+2}$, N$^{+3}$ and N$^{+4}$  do not set useful constraints on the abundances of these ions in any of our targets. Therefore ICF(N) is derived from Eq.~A2 of \citet{KB94}. Note that ICF(N) is uncomfortably large for MG~13 and SMP~23. Therefore only a lower limit for N is given in Table~\ref{tab:Abund} for MG~13 based on the N$^+$ abundance alone (although N/O can still be derived from N$^+$/O$^+$); for SMP~23 the N abundance is especially uncertain, being estimated from $<1$\% of the total, although no recombination lines of \ion{N}{3} were detected. 

Two ionization stages of neon (Ne$^{+2}$ and Ne$^{+3}$) are observable in the optical, while four stages (Ne$^+$, Ne$^{+2}$, Ne$^{+4}$, and Ne$^{+5}$) are observable in the IR. Most of these lines are faint, however, and in many cases only upper limits are available for stages other than Ne$^{+2}$. Following \citet{KB94}, we assume the Ne$^+$ contribution is negligible, and set ICF(Ne)=1.5 when Ne$^{+4}$ is observed but Ne$^{+3}$ is not; and use their Eq.~A28 if only Ne$^{+2}$ is observed. In all cases the ionic abundances from the IR lines are used where available, a point we discuss in \S4. 
We also note that the upper limits for undetected ions are all consistent with the adopted ICFs. Curiously, the abundances of Ne$^{+2}$ as derived from the IR lines are roughly 50\% greater than those derived from the optical lines, the latter being derived entirely from [\ion{Ne}{3}] $\lambda3868.7$ emission. We discuss the fidelity of the optical measurement in \S4.1, but note here that \citet{PB99} found a similar effect, though with a much larger discrepancy between the optical and IR. 

As with oxygen, the first two ionization stages of sulfur (S$^+$ and S$^{+2}$) are observable in the optical spectra for all of the PNe in our sample, although S$^{+2}$ is not actually detected in two of the objects. In the IR spectra, S$^{+2}$ is detected in six objects, and useful upper limits are established for an additional five; S$^{+3}$ is detected in nine objects, and useful upper limits are established for two others. A comparison of the S$^{+2}$ abundance from the optical and the IR shows generally good agreement, and upper limits in the IR equal or exceed those derived from the optical lines. Once again we adopt the S$^{+2}$ abundance from the IR lines where available. 
The S$^{+3}$ abundances and upper limits shed light on the computed ICFs, which we derived from Eq.~4$e$ of \citet{KH01} where necessary. For five objects (SMP~8, SMP~11, SMP~13, SMP~24, and SMP~27) the abundance from the simple sum of the first three ionization stages (which we adopt, setting ICF(S)=1.0) is within $\sim11$\% of that derived from (S$^{+}+$S$^{+2}) \times$ICF(S). For SMP~18 the upper limit to S$^{+3}$ shows that the ICF is within 10\% of the correct value. Of the remaining four objects, three of them (SMP~14, SMP~17, and SMP~19) have relatively high ionization, with the abundance of S$^{+3}$ exceeding that of S$^{+2}$ and suggesting that contributions to the total S abundance from even higher ionization stages may be significant. In these cases, the S abundance derived from ICF(S) is even lower than the S abundance derived from the simple sum of the observed ions; we adopt the simple sum as a lower limit. 
Finally, S$^{+2}$ is not detected in the optical spectrum of SMP~23 and only an upper limit is available from the IR spectrum, yet S$^{+3}$ exceeds S$^+$ by an order of magnitude. In this case we adopt the straight sum of the abundances from the observed ions as a lower limit. 

All of the relevant ionization stages of Ar are observable in the optical, except for Ar$^+$ which is only observable in the IR. The \textit{Spitzer}/IRS spectrograms were too shallow to detect the [\ion{Ar}{2}] 6.99~$\mu$m line, so only upper limits to Ar$^+$ could be computed. In all cases these limits are much too large to be useful for constraining the derived ICF(Ar). Therefore the prescription for ICF(Ar) is taken from Eq.\ A30 of \citet{KB94} to correct for its absence. Ar$^{+2}$ is also observed in the IR, but only upper limits are available for these targets; the limits are all consistent with the ionic abundances determined from the optical spectra. 

\subsubsection {Uncertainties}

We computed the uncertainties for each element abundance by determining the effect of the uncertainty in $T_e$, and adding in quadrature the effect of uncertainties in the emission line intensities that were quantified in \S2. 
The uncertainty for abundances determined solely from optical emission lines is $\sim0.10$ dex, owing to the strong dependence of abundance on $T_e$, while for the IR lines it is $\sim0.05$ dex. Thus, the magnitude of the uncertainties in the elemental abundances varies considerably, depending upon the wavelengths of the transitions for the dominant ions. Broadly speaking the uncertainties are 0.1 dex or better, but may be higher in individual cases as noted above. No allowance has been made for the statistical or (the likely larger) systematic uncertainty in the derived ICFs, but since the corrections are generally of modest size (except for N) these uncertainties would not dominate the errors. Generally speaking, the magnitude of our adopted ICFs is comparable to those in the literature \cite[see, e.g.,][]{KB94, KH01,HKB04} in those cases where only optical emission lines are available for analysis. The main exception is for MG~8, which will be discussed separately below. The relatively large ICF(N) for most of our objects results from the availability of only one observable ionization stage (N$^+$) in the optical and IR, the moderate- to high-ionization of most of this sample, and from the relatively low enrichment of N in these non-Type~I PNe. 
 
\subsection {Comparison to Published Abundances}

Our sample of objects presents an unusual opportunity to examine in detail the sources of uncertainty in the determination of elemental abundances using direct methods, and to explore the discrepancies between different studies. All of the objects in this sample have been observed by at least two, and usually a few observers. They are all angularly small, so the entire nebula was most likely included in the spectrograph aperture, and therefore ionization stratification and differential effects of dust are not factors in the comparison. All of these objects have been observed with at least one space-based spectrograph: most have been observed with both \textit{HST}/STIS and \textit{Spitzer}/IRS. Thus, the absolute fluxes of key emission lines, and the relative fluxes of many other lines over a broad wavelength range, are well determined. 
It is especially useful to compare abundances that are derived from optical lines to those derived from the IR, both as a check of consistency and as a means of evaluating the veracity of the ICFs. The advantages of deriving abundances from the IR fine structure lines have been enumerated before \citep[see, e.g.,][]{BernardSalas01}. Briefly, the IR emission lines are stronger than the collisionally excited emission lines in the optical, they are not greatly affected by interstellar extinction, the abundances derived from these lines are little affected by uncertainties in $T_e$, and more ionization stages can be observed than are available in the optical alone. To this we would add that the IR emission line strengths measured from \textit{Spitzer}/IRS spectra are more accurately calibrated than those from most ground-based spectra (optical or IR). In \S3.3.1 we compared the elemental abundances as determined from optical emission lines and ICFs to those derived from a combination of optical and IR lines, without ICFs. This idea is not new and appears to have been first suggested by \citet{PB99}, but here we are able to make comparisons for a greater number of ICF prescriptions. 

\begin {figure}
\epsscale{0.4}
\plotone {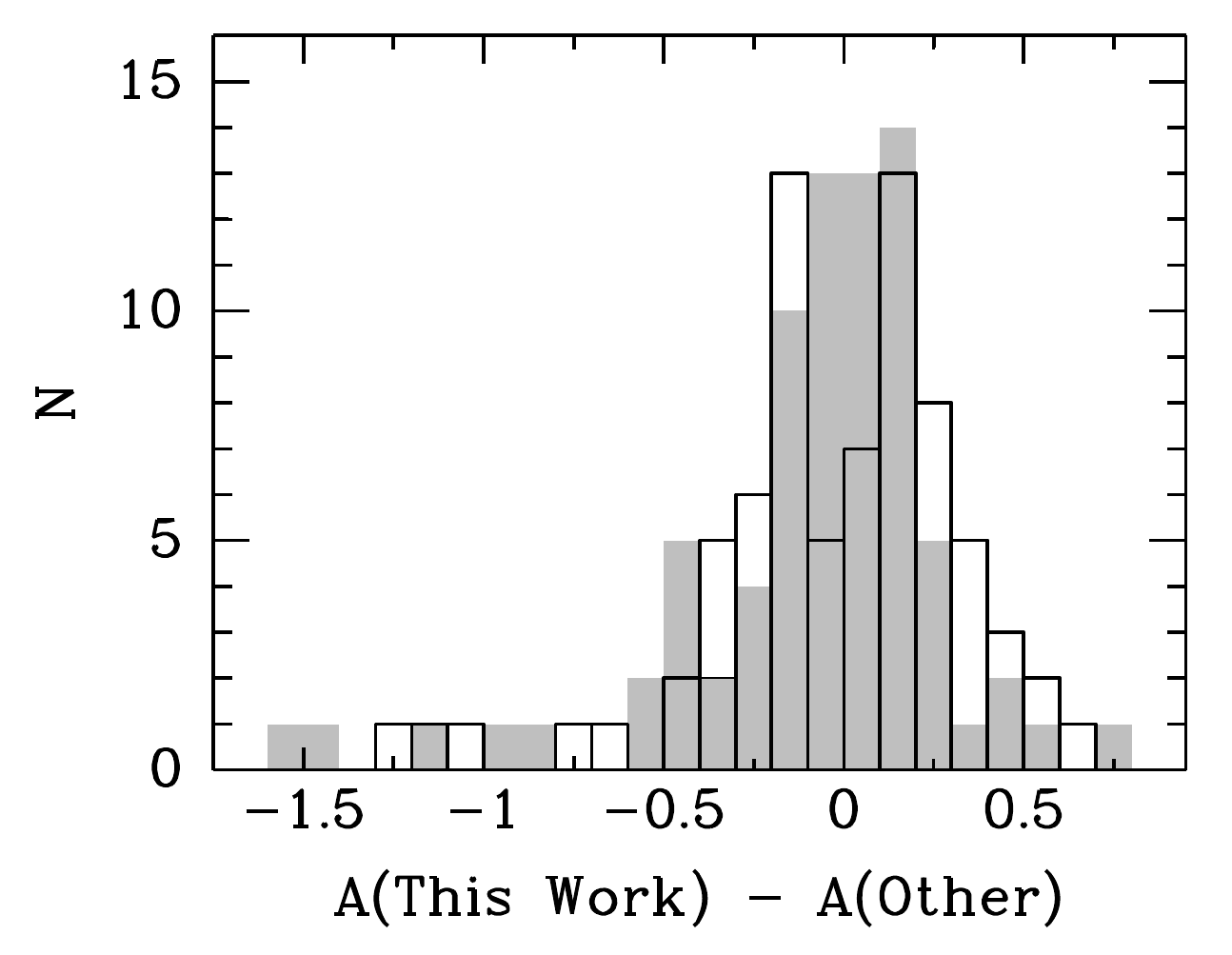}
\figcaption[f2.pdf]{ 
Discrepancies in elemental abundances between the results of this paper and those of LD06 (\textit{open boxes}) and IMC07 (\textit{filled boxes}). Plot shows histograms of the differences in the log abundance over all elements considered here.  
\label{fig:diffAB}}
\end {figure}

We compared our elemental abundances in Table~\ref{tab:Abund} to those presented in the large compilations of LD06 and IMC07 for the same objects. The last four rows of the Table give the median abundance (excluding limits) for targets in this sample, the median abundances from LD06 and from IMC07 for targets in common (excluding highly uncertain abundances), and the mean of \ion{H}{2} regions in the SMC from \citet{Den89}. 
Figure~\ref{fig:diffAB} shows histograms of the differences in A(X), the logarithm of the abundance for element X, over all elements. The resulting distributions are quite broad: the half-width exceeds 0.15~dex and large wings extend to 1.5~dex. It is also useful to compare the differences in the median abundance for each element between the two studies, which we give in the last few rows of Table~\ref{tab:Abund}. 
At the outset we expected an offset of up to +0.07~dex and $-0.03$~dex (for LD06 and IMC abundances, respectively) derived from faint lines because of the flux discrepancy identified in \S2.2. In fact, we find our abundances for He are higher by 0.16~dex. While the origin of this discrepancy is not entirely clear, part of the reason may be our use of the more recent calculation of the \ion{He}{1} recombination coefficients in the Case B limit by \citet{Porter_etal05}. 
For N, our median abundance is lower than those of LD06 by 0.34~dex, although if the values noted by LD06 as poorly determined are excluded, then our median abundance \textit{exceeds} that of LD06 by only 0.14~dex for PNe in common, which is within their quoted uncertainty. Our median abundance is less than than that of IMC07 by 0.09~dex, which is within the quoted uncertainties. We believe our N abundances are the more accurate, owing mostly to improved accuracy of the emission line intensities. 
Our median O abundance is lower by 0.07~dex and 0.06~dex than that of LD06 and IMC07, respectively, which is within the quoted uncertainties in both studies. However, this similarity belies the much larger range of values, which will be discussed in \S4.2. 

Our median Ne abundance is higher by 0.22~dex than that of LD06, but this is almost entirely due to our use of abundances from the IR lines; had we used only the optical lines, the difference would have been near +0.04~dex. Our median Ne abundance agrees well with that of IMC07, to within 0.05~dex. 
It is not clear why the abundances from the optical and IR emission lines are so discrepant for this element. Two potential sources of error, the flux calibration and the extinction constant, were validated by the consistency of the ratios of [\ion{O}{2}] $I(\lambda3727)/I(\lambda7325)$ and [\ion{S}{2}] $I(\lambda4072)/I(\lambda6723)$ with the adopted extinction constants and $T_e$, and by the consistency of the ionic abundances as derived from the optical and IR for other elements. Since a similar discrepancy was noted by \citet{PB99}, we suspect the problem may lie with the supporting atomic data. 
Our median S abundance is 0.66~dex and 0.61~dex lower than LD06 and IMC07, respectively. Although this difference is exceptionally large, in the first case it is based on only the 3 objects in common that were not excluded from the LD06 study as unreliable. Furthermore, the [\ion{S}{2}] line fluxes are weak: the strongest in this sample are $\sim10$\% of H$\beta$, and the weakest are $\sim1$\%. Thus many of the S abundances are subject to the systematic errors of the type described in \S2.2. 
We are confident in our S abundances because the combined optical and IR spectra cover most ionization stages of S, and that ICF(S) as computed from the optical emission lines is in excellent agreement with that inferred from the abundances as computed from both optical and IR lines. 
Finally, our median Ar abundance is only 0.08~dex and 0.06~dex lower than that of LD06 and IMC07, respectively, and well within the quoted uncertainties for this element. 

\section {Discussion}

\subsection {Errors in Elemental Abundances}

The uncertainty in the determination of the elemental abundances using direct methods depends upon a variety of primary and secondary factors, including the uncertainty in the observed emission line intensities, uncertainties in the extinction, $T_e$ and $N_e$, and in the determination of the ICFs, which themselves depend upon abundances of other ionic species. It has often been argued that the majority of abundance uncertainties are predominantly statistical in nature, and are governed by shot noise in the emission line intensities. While it is true that some ionic abundances are derived from very weak emission lines, the majority are derived from lines with intensities that are typically greater than 10\% that of H$\beta$, with corresponding quoted uncertainties of order $\sim10$\%. Yet Figure~\ref{fig:diffAB} shows that the abundance discrepancies found here are $\sim0.15$~dex (40\%) or more. We argue that the magnitude of the discrepancies identified here includes a significant component of \textit{systematic} errors in emission line intensities, which when propagated through the analysis result in greatly magnified errors in elemental abundance determinations. 

\begin {figure}
\epsscale{0.5}
\plotone {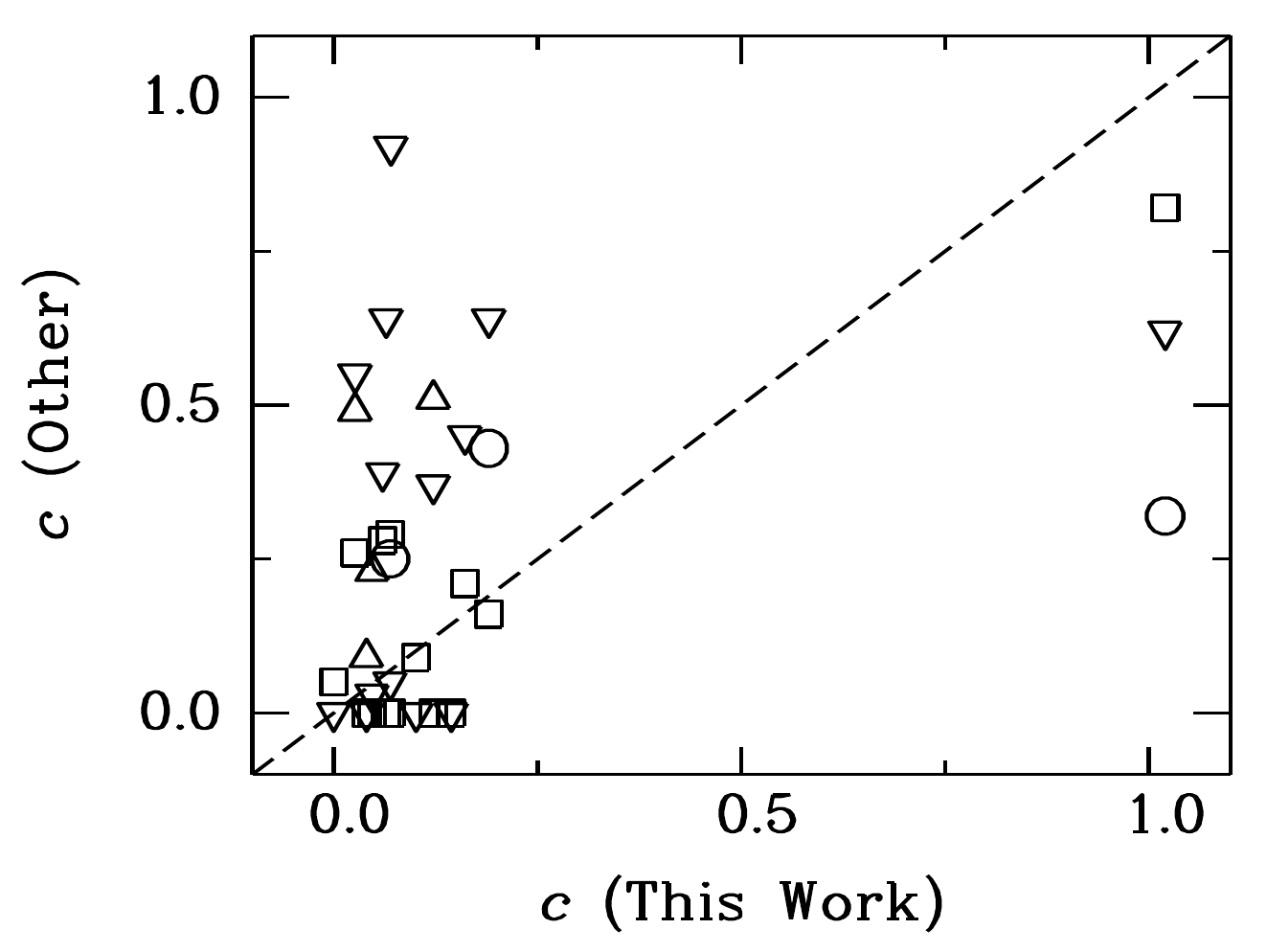}
\figcaption[f3.pdf]{ 
Adopted extinction constant from this paper vs. those of \citet{MBC88}, \citet{MD91a, MD91b}, \citet{LD_96, LD_06}, \citet{IMC07}, with 1:1 line indicated (\textit{dashed line}). Symbols as in Fig.~\ref{FlxCmp}. 
\label{fig:diffC}}
\end {figure}

Systematic errors in emission line intensities can have a variety of causes, including flawed observing technique and poor flux calibration. If, for example, a ground-based spectrograph is configured to obtain a large wavelength coverage with a narrow slit, the resulting observations of compact targets may be compromised by color-dependent slit losses due to atmospheric dispersion even at modest airmass \citep{Filippenko82} unless care is taken to mitigate the effect during the observation. The effect of such slit losses, which are often most severe at the blue end of the spectral range, can be manifested in inferred interstellar extinction constants that are systematically too large. Figure~\ref{fig:diffC} compares the extinction constants (expressed as $c$, the logarithmic extinction at H$\beta$) used here with those measured by \citet{MBC88}, \citet{MD91a, MD91b}, \citet{LD_96, LD_06}, and IMC07\footnote{In the end, IMC07 adopted the mean of extinction constants from several sources.}.
Errors of this sort, or in flux calibration, can be detected using sanity checks with emission lines that span a large range of wavelength, such as ensuring that the [\ion{O}{2}] $I$($\lambda3727$)]/$I$($\lambda7325$) and [\ion{S}{2}] $I$($\lambda4072$)]/$I$($\lambda6725$) ratios are consistent with the adopted $c$ and $T_e$. 
Generally, errors in $c$ smaller than 0.2 do not cause large errors in ionic abundances derived entirely within the optical band, but errors this large may result in substantial discrepancies between ionic abundances derived from UV and IR emission lines.  Errors in the extinction constant are unlikely to cause significant errors in the determination of $T_e$, as the most commonly used ratios (from \ion{O}{3} $I$($\lambda5006.8$)/$I$($\lambda4363.2$) and [\ion{N}{2}] [$I$($I$($\lambda6583.4$)]/$I$($\lambda5754.6$) span relatively modest wavelength baselines. 

\begin {figure}
\epsscale{0.5}
\plotone {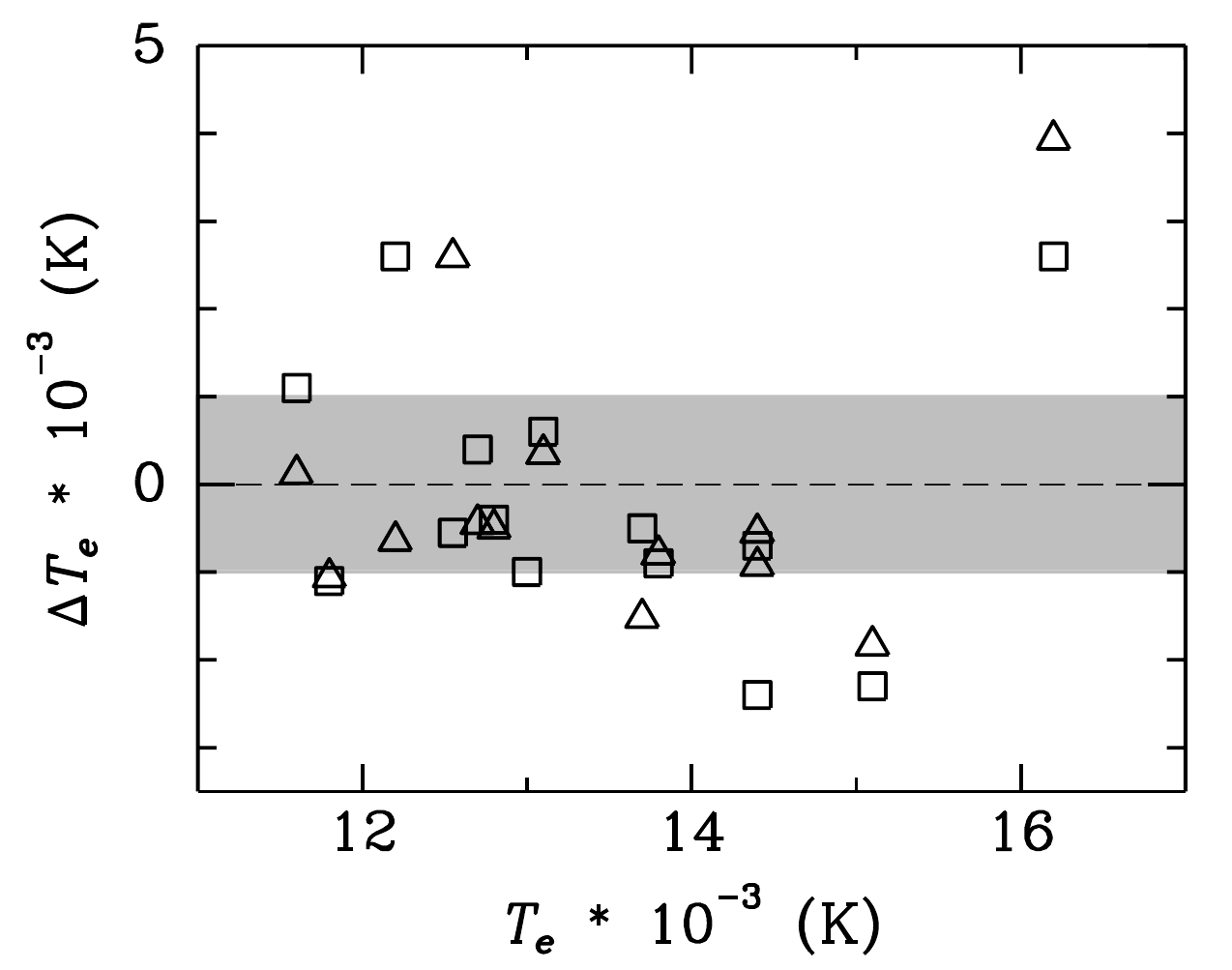}
\figcaption[f4.pdf]{ 
Adopted $T_e$ (in 1000 K) from this paper vs. the difference in $T_e$ between those of other authors (LD06; and IMC07) and this paper. Symbols as in Fig.~\ref{FlxCmp}. Differences exceeding 1000~K (i.e., symbols lying outside the \textit{grey box}) can result in differences in inferred abundance exceeding 0.10~dex. 
\label{fig:Te}}
\end {figure}

The ionic abundances depend linearly on the observed emission line intensities. However, the dependency on $T_e$ is either minor (for transitions in the IR) or exponential (for optical and UV transitions). It is this exponential dependence of the optical emission lines on $T_e$ that can inflate modest systematic errors in fluxes to large errors in ionic abundances. Systematic errors of only 10\% in emission line intensities can lead to errors in $T_e$ of $\sim1000$~K, depending on the diagnostic used. For temperatures in the range considered here, an error of 1000~K can translate to an error in ionic abundances as determined from optical emission lines of $\sim0.08$--0.15~dex. Judging by the magnitude of the discrepancies between $T_e$ as determined here and those of LD06 and 
IMC07\footnote{The largest apparent discrepancy in $T_e$ was for MG~8, where IMC07 derived 23,900~K compared to our 13,000~K. However, they have adopted an intensity of 5\% of H$\beta$ for the auroral emission line of [\ion{O}{3}] $\lambda4363$, which appears to have originated from LD06. This emission line was in fact listed by LD06 as an \textit{upper limit} to the intensity from the spectrum published by \citet{Vass_etal92}. The ratio of this intensity to the intensities of the nebular [\ion{O}{3}] lines from \citet{Vass_etal92} would yield the erroneous $T_e$ that was used by MC07. Since this appears to be an error, rather than a difference in measurement or technique, we do not include that high $T_e$ in Fig.~\ref{fig:Te}}, 
shown in Figure~\ref{fig:Te}, it is not surprising that the elemental abundance discrepancies can be much larger than the quoted statistical errors for the individual line fluxes. Other factors can inflate abundance discrepancies as well, such as errors in the supporting atomic data and the adopted ICF recipes. The former can lead to discrepancies of 30\% if the latest atomic data are not used, while in practice there is considerable agreement on the ICF recipes.  

\subsection {Abundances as Tracers of Prior Stellar Evolution}

The small number of objects in this sample of PNe makes it difficult to gain deep insight into nucleosynthesis and dredge-up processes of AGB stars in the SMC, but a few important conclusions can be drawn. \citet{Villa_etal04} determined the central star masses of 10 of the PNe in this sample, and with two exceptions they are $\approx0.59$~M$_\sun$, implying moderate- to low-mass progenitors. This inference is reinforced by the PN abundances: 
Figure~\ref{fig:NOvsHe} shows N/O vs.\ He for our sample of SMC PNe, along with the data from LD06 and IMC07 for the same objects. This diagram has traditionally been used as a diagnostic to discriminate between Type~I and non-Type~I MCPNe \citep{TPP97}. Based on morphological and kinematic data, Type~I PNe in the Galaxy are associated with bipolar shapes and higher mass progenitors. Chemically, they show substantial enrichments of N and lesser enrichments of C relative to O, which is consistent with theoretical predictions for hot-bottom burning (HBB) in the most massive PN progenitors at the end of AGB evolution \citep[see, e.g.,][]{Marigo_etal03}. Non-Type~I PNe show lesser enrichments in N/O and He, substantial enrichments of C, and typically have round or elliptical morphologies. 
There are clearly no Type~I PNe in our sample, although all of the nebulae except SMP~11 show high N/O and high He and N abundance relative to \ion{H}{2} regions. This is consistent both with the nebular morphology (again, except for SMP~11), and with the high abundance of C in the 9 objects where it has been measured \citep{LD_96, Stang_etal09}. 
MG~8 is an especially interesting object in this context in that \citet{Villa_etal04} derived a central star mass of 0.88~M$_\sun$, implying a progenitor Main Sequence mass of at least 3.5~M$_\sun$ according to the models of \citet{VW94}. This is close to the minimum mass that is thought (at SMC metallicities) to be required for hot-bottom burning (HBB) during the prior AGB evolution, which would have converted C to N very efficiently. The comparatively modest N abundance in MG~8 sets a useful constraint on the maximum progenitor mass in the SMC that still does not undergo HBB during AGB evolution. 

Further insight into nucleosynthetic processing in the progenitor AGB stars can be gained from Figure~\ref{fig:NOvsO}, which shows N/O vs.\ O in the nebular gas, and compares our data to that from LD06 and IMC07 for this sample. The rather large scatter in the LD06 and IMC07 data for both N/O and O is greatly reduced in our data, which again illustrates rather dramatically the benefit of improved accuracy in abundance determinations. If SMP~11 is excluded, the points are consistent with no decrease in N/O with increasing O. Although our sample is rather small, stellar evolution models \citep{Marigo_etal03} predict that the O-N cycle will not be activated for AGB stars that do not experience HBB, which is consistent with the lack of a significant trend (i.e., zero slope) in Figure~\ref{fig:NOvsO} for the progenitor AGB stars in this mass range. 

Planetary nebulae can also probe stellar evolution theory for massive stars \citep[see, e.g.,][]{HKB04} by studying the abundances of elements such as Ne, S, and Ar that are not predicted to be altered during the course of prior AGB evolution; these abundances would therefore reflect that of the interstellar medium when and where the progenitor star was formed. Very often, these elements are compared to the abundance of O which is also assumed to be unaltered for stars in this mass range. Figure~\ref{fig:NEvsO} shows the abundance of Ne vs.\ O, which is well fit with the relation log~(Ne) $= -1.96 + 1.16$ log~(O), with an RMS of 0.11 when MG~8 is excluded. This fit is in nearly perfect agreement with that of \citet{HKB04}. The relation also agrees with the results of \citet{Stang_etal06} for Galactic PNe, also shown in Fig.~\ref{fig:NEvsO}, in the sense that their trend line would intercept the current sample of SMC PNe which have a lower average O abundance of roughly half a dex. 
The tight correlation is strong evidence that Ne and O vary in lock-step, as argued by several investigators \citep{HKB04, RMc06, MCI09}, and that therefore O is basically unaltered during the course of evolution of stars that produce PNe. 

Concerning S, we note that log(S/O)$=-1.98$ (in the median) for this sample, which is lower by $\sim0.3$~dex compared to that for SMC \ion{H}{2} regions (see Figure~\ref{fig:SvsO}). This is consistent with the ``S anomaly'' identified by \citet{HKB04} in Galactic PNe, who suggested that it most likely results from erroneous S abundance determinations owing to an inadequate accounting of unobserved S$^{+3}$ in the ICF(S). But since our S abundances are derived mostly from mid-IR emission lines, S$^{+3}$ is detected if it is present in significant quantity. Thus, the S anomaly appears to be genuine, and persists even in a low metallicity environment. \citet{HKB04} also suggested a similar deficit for Ar, and we find log(Ar/O)$=-2.47$ (in the median) for our sample. This value is also lower by $\sim0.3$~dex compared to that for SMC \ion{H}{2} regions. 
The case for an Ar deficit is a bit less convincing (see Figure~\ref{fig:ARvsO}) since we rarely detected the very faint Ar$^{+3}$ and Ar$^{+4}$ emission lines in the optical spectra. One possible explanation may be that the abundances of S and Ar in the interstellar medium of the SMC have changed relative to O and Ne over the lifetimes of the progenitors of these PNe. However, we would caution that the source of the \ion{H}{2} region abundances \citep{Den89} is some twenty years old, and that a study of a larger sample with greater depth, better and more complete wavelength coverage, and updated atomic data is perhaps overdue. 

\subsection {Unusual Planetary Nebulae}

MG~8 is a very low excitation nebula, but its slightly elliptical morphology leaves little doubt of its classification as a PN \citep{Shaw_etal06}, although it could be a bipolar viewed pole-on. Its spectrum shows broad stellar emission from \ion{He}{2} and \ion{C}{4}, indicating the presence of a wind. It is unusual in that the abundance of Ne is low by about an order of magnitude for its O abundance (see Fig.~\ref{fig:NEvsO}), yet the Ar abundance is high by about 0.5~dex (see Fig.~\ref{fig:ARvsO}) compared to the other PNe in this sample. We note, however, that the uncertainty in the O and Ne abundances is relatively high for this object, at 0.19 and 0.18 dex, respectively. 
The ICF(Ne) is higher for MG~8 than the other PNe, but the absence of nebular He$^{+2}$ and the modest level of O$^{+2}$/O suggests that an even higher ICF(Ne) is not warranted. \citet{Villa_etal04} determined for this object a central star mass of 0.88~M$_\sun$, T$_{eff}= 66,500$~K, and log~L/L$_{\sun} = 4.33$. 
Although the high mass raises the possibility of a higher efficiency for depletion of Ne via $^{22}$Ne($\alpha,n$)$^{25}$Mg processing in the progenitor AGB star, it is difficult to see how it could result in an order of magnitude deficit. The abundances of the other elements in MG~8 are typical for the PNe in this sample, although no C abundance is available. 

SMP~11 was classified by \citet{Stang_etal03} as bipolar, with a significantly larger extinction than most SMC PNe. However, its chemical composition is quite typical of the average of \ion{H}{2} regions in the SMC \citep{Den89}. More specifically, N is not elevated as it is for the other PNe, and certainly not as much as one would expect for a bipolar morphology. Given that its morphology and excitation is not inconsistent with a compact \ion{H}{2} region, its earlier classification as a PN is in some doubt. 

\section {Conclusions and Future Work}

We have presented new emission line fluxes for several SMC PNe based on deep, high resolution optical spectra and low-resolution mid-IR spectra. We have used a variety of techniques to assure the accurate calibration of our measurements, and have endeavored to determine meaningful statistical uncertainties in the derived elemental abundances. Our optical fluxes show excellent agreement with extant space-based fluxes, but the agreement with prior ground-based studies shows significant, systematic discrepancies especially at faint intensity levels. Such discrepancies propagate to differences in quantities derived from ratios of emission line fluxes, such as the extinction constant, $T_e$, ionic abundances, and ICFs, the cumulative effect of which can result in comparatively large discrepancies in derived elemental abundances among authors. 

We derived ionic abundances using IR fine-structure emission lines where possible, which have the advantages of high emissivity relative to optical lines, and insensitivity to uncertainties in the derived $T_e$ and extinction. Combined with the superior calibration of the space-based IR data compared to ground-based data, the derived ionic abundances have substantially lower uncertainty. The IR emission lines are important for deriving ionic abundances for several ions that are not observable in the optical. They also provide an important check of the ionic abundances derived from the optical data for ions in common, where the agreement is very good except for Ne$^{+2}$. 
We were able to validate elemental abundances derived from the combination of optical emission lines and traditional ICFs with those derived from a combination of optical and IR lines (or limits thereto). We found excellent agreement for O, S, and Ar; the poor agreement for Ne is due to a consistent offset in the Ne$^{+2}$ abundance, which may point to a problem with the supporting atomic data. In the end, our data have significantly improved the accuracy of the elemental abundances for Ne and S in the target nebulae, with considerable improvement in the abundances of O and Ar as well, as compared to those in the literature. 

We have explored evidence of nucleosynthetic processes in the progenitor AGB stars, and we confirm the expectations from stellar evolution theory that the low-mass PN central stars in our sample show no signs that they underwent HBB on the AGB, nor is there evidence of destruction of O through the O-N cycle. 
One object in this sample, MG~8, was evidently born of an intermediate mass progenitor shows peculiar abundances
of Ne and Ar. Given its lack of significant N enrichment and inferred progenitor mass of 3.5~M$_\sun$, this object sets a useful constraint on the maximum progenitor mass at this metallicity that does not undergo HBB during AGB evolution. 
We confirmed that the Ne and O abundances vary in lock-step, join smoothly with the relationship seen in Galactic PNe, and likely are not affected by prior AGB evolution. We also showed that the S anomaly (i.e., low S/O relative to local \ion{H}{2} regions) seen in Galactic PNe appears to be genuine, and suggest that such an anomaly might also be present for Ar at least for our SMC sample. 
Finally, we noted that the abundances of SMP~11 closely match those of SMC \ion{H}{2} regions, which suggests it may not be a PN after all. 

\begin {figure}
\epsscale{0.65}
\plotone {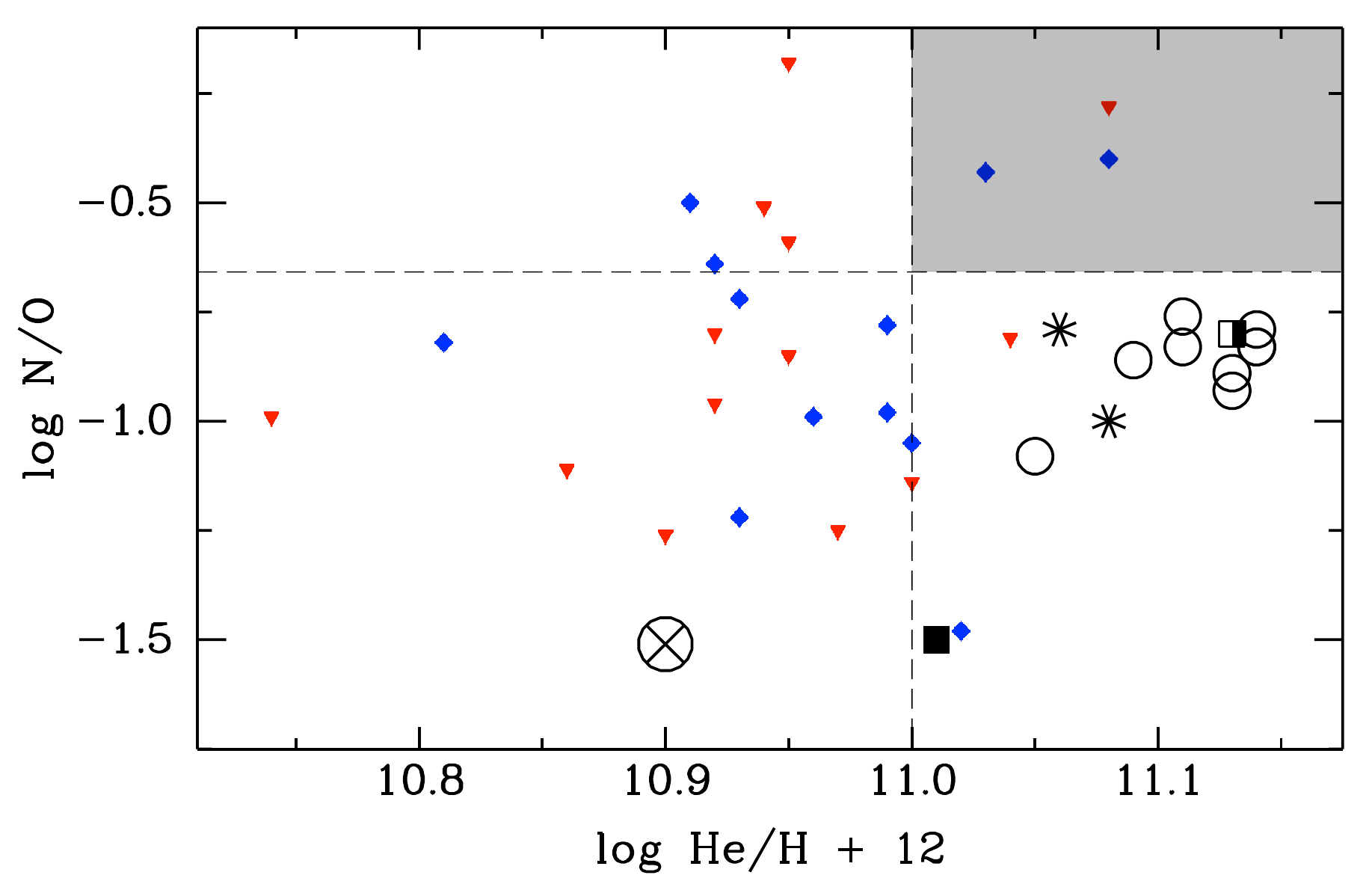}
\figcaption[f5.pdf]{ 
N/O vs.\ He/H for this sample of SMC PNe, with the domain of Type~I nebulae indicated (\textit{shaded region}) as revised for SMC abundances by \citet{TPP97} and with restrictions adopted from LD06. Symbols indicate morphological type: Round (\textit{circles}); Elliptical (\textit{asterisks}); bipolar (\textit{squares}); bipolar-core (\textit{half-filled squares}). Also plotted are data from LD06 (\textit{small inverted triangles}) and IMC07 (\textit{small diamonds}) for the same objects. 
Average abundances of SMC \ion{H}{2} regions \citep{Den89} is indicated (\textit{circled cross}). 
[\textit{See the electronic edition of the Journal for a color version of this figure.}]
\label{fig:NOvsHe}}
\end {figure}

\begin {figure}
\epsscale{0.65}
\plotone {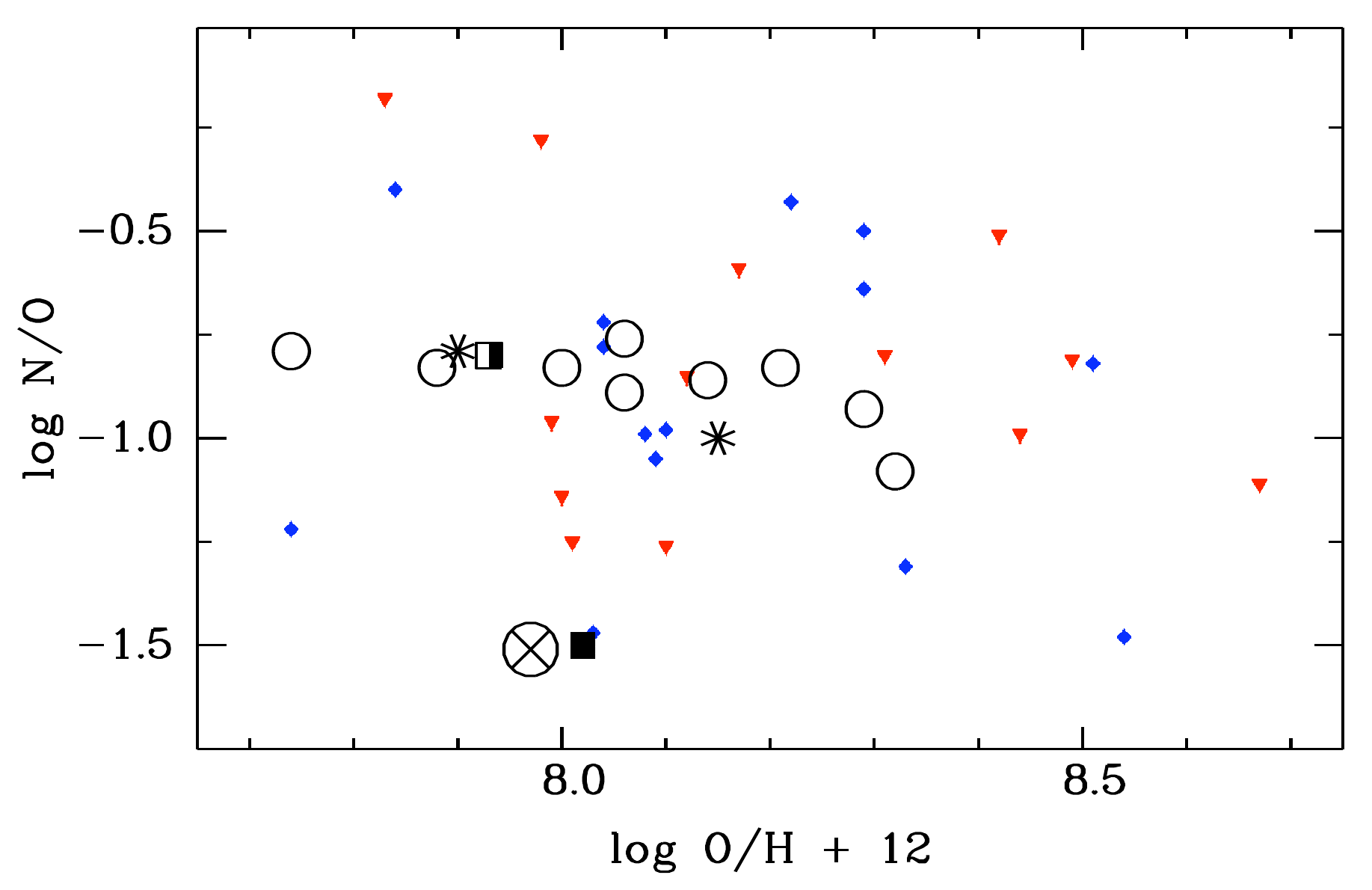}
\figcaption[f6.pdf]{ 
N/O vs.\ O/H for this sample of SMC PNe, along with data for the same objects from LD06 and IMC07. Symbols as in Fig.~\ref{fig:NOvsHe}. 
[\textit{See the electronic edition of the Journal for a color version of this figure.}]
\label{fig:NOvsO}}
\end {figure}

\begin {figure}
\epsscale{0.65}
\plotone {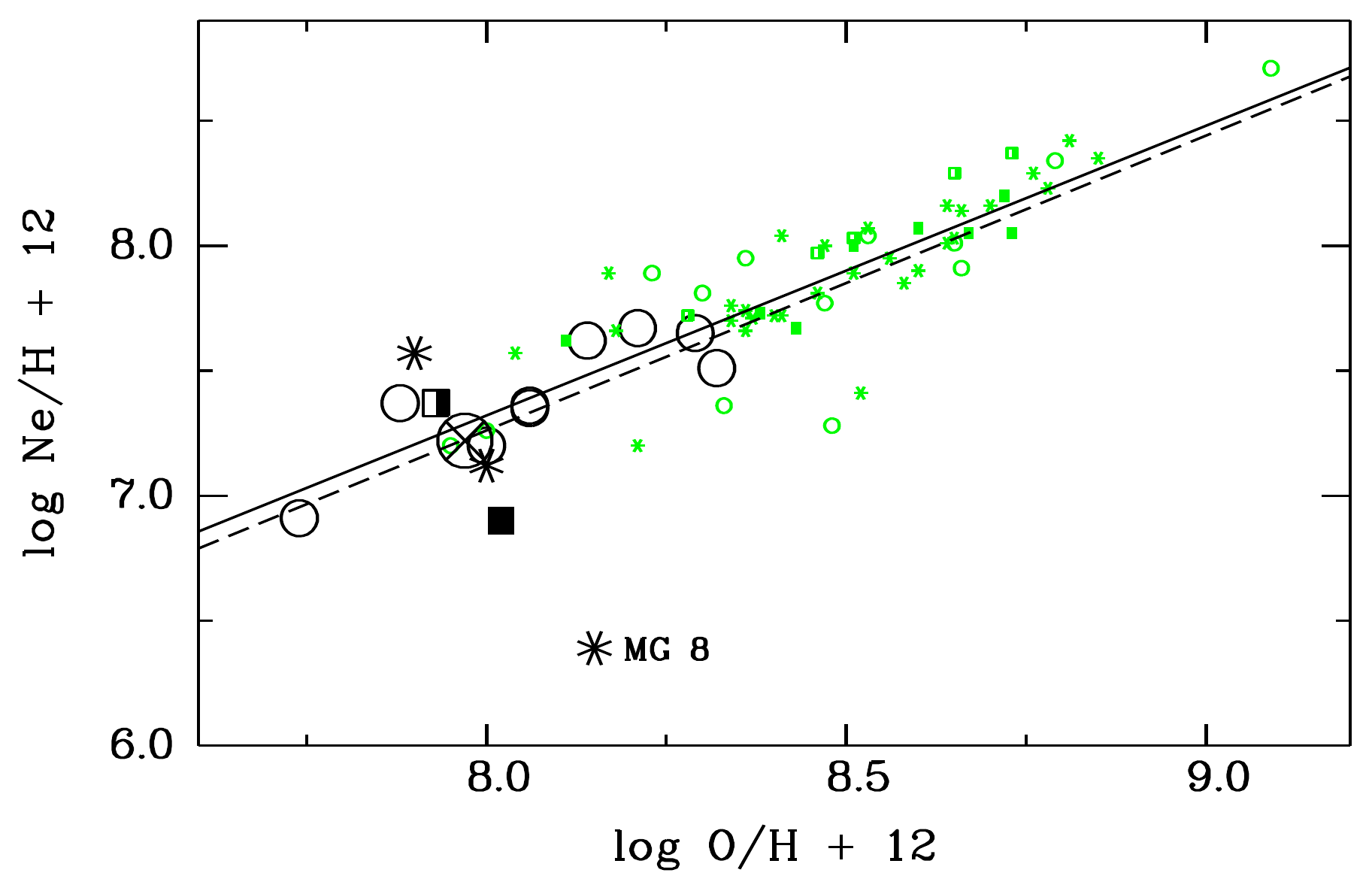}
\figcaption[f7.pdf]{ 
Ne/H vs.\ O/H for this sample of SMC PNe (\textit{large symbols}) and for the Galactic PN sample of \citet{Stang_etal06} (\textit{small green symbols}). Best-fit relation (\textit{solid line}) agrees very well with that from \citet{HKB04} (\textit{dashed line}) for Galactic PNe. Symbol types as in Fig.~\ref{fig:NOvsHe}. 
[\textit{See the electronic edition of the Journal for a color version of this figure.}]
\label{fig:NEvsO}}
\end {figure}

\begin {figure}
\epsscale{0.65}
\plotone {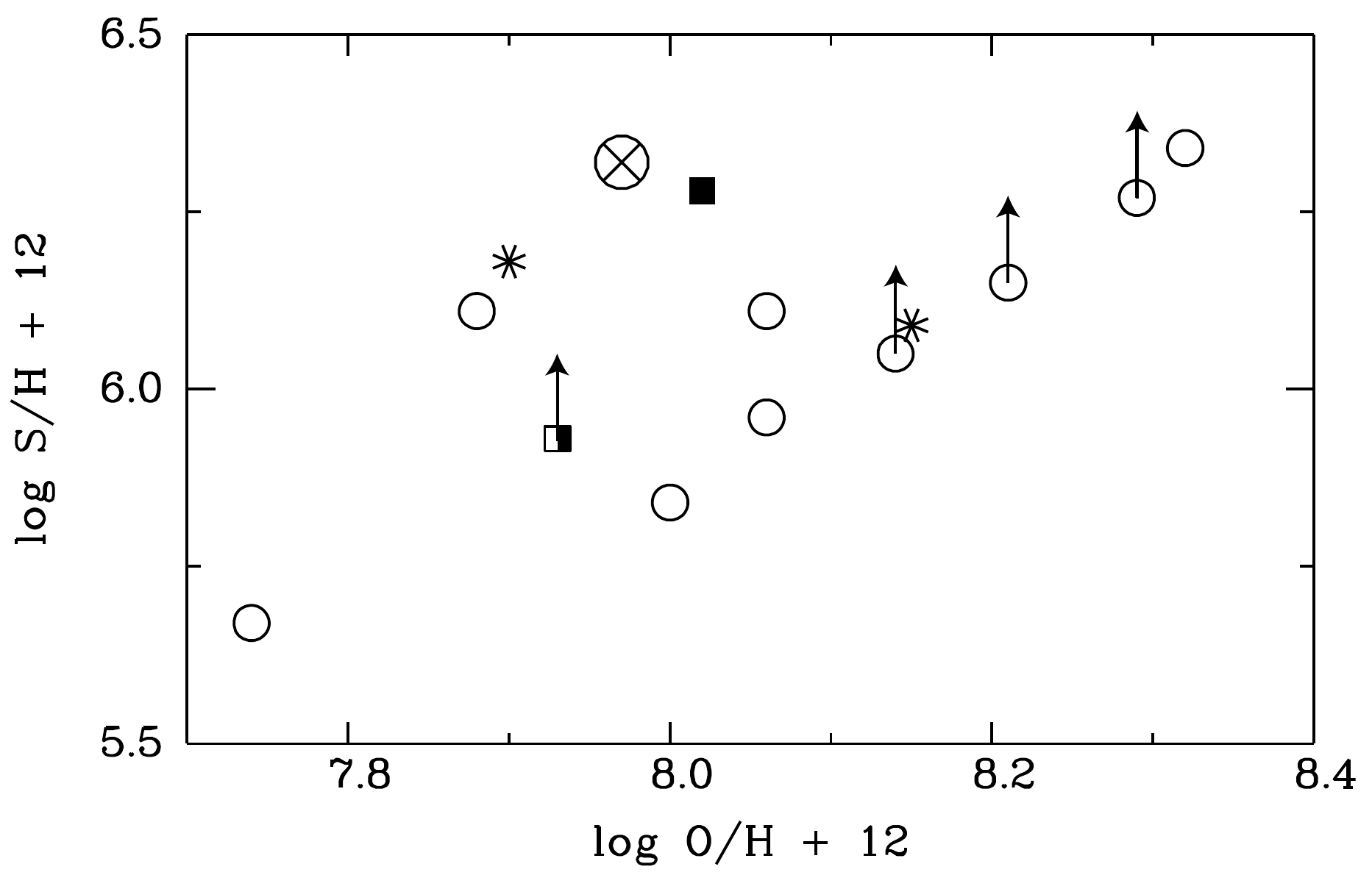}
\figcaption[f8.pdf]{ 
S/H vs.\ O/H for this sample of SMC PNe. Symbols as in Fig.~\ref{fig:NOvsHe}. 
\label{fig:SvsO}}
\end {figure}

\begin {figure}
\epsscale{0.65}
\plotone {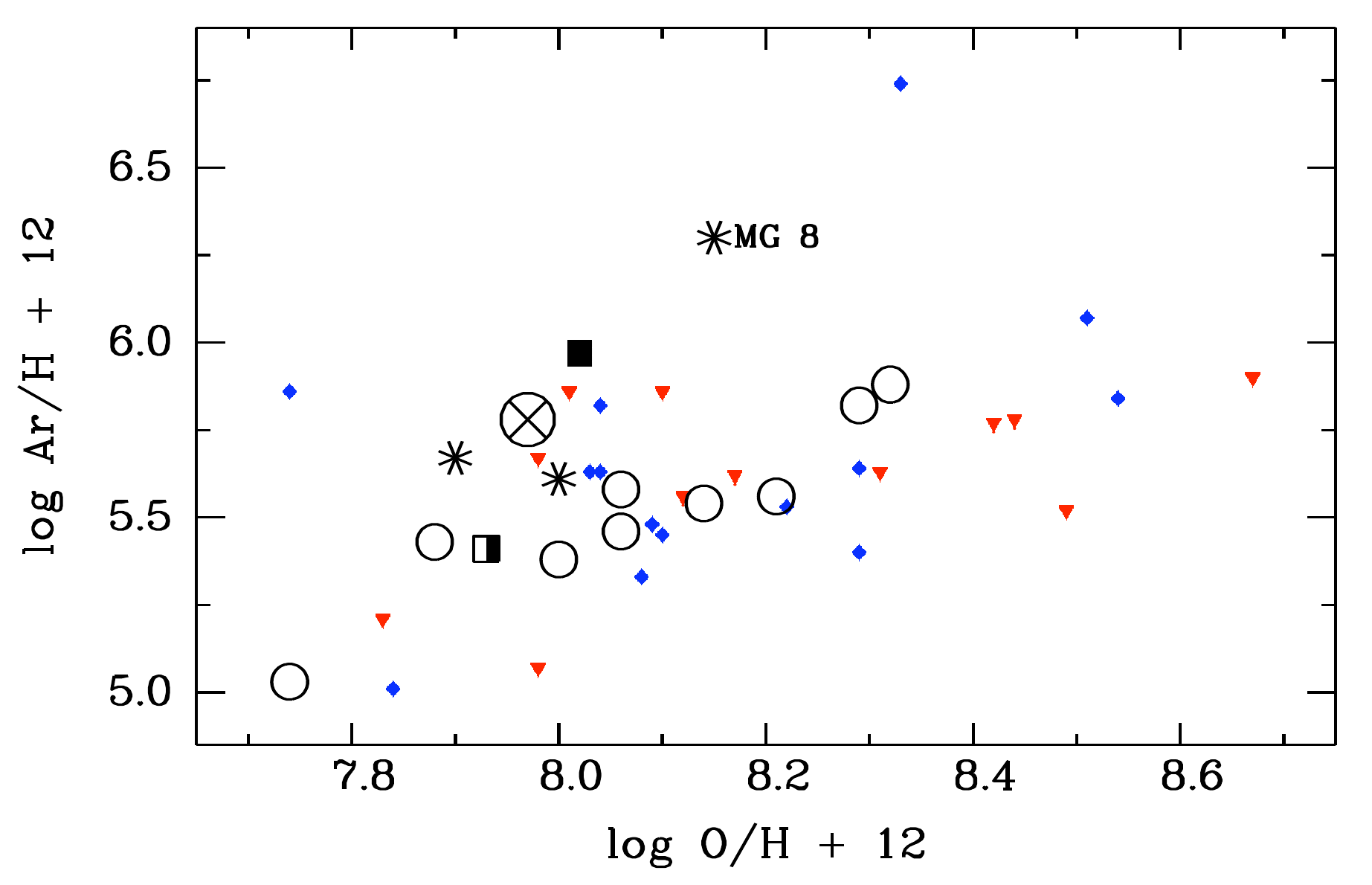}
\figcaption[f9.pdf]{ 
Ar/H vs.\ O/H for this sample of SMC PNe, along with data for the same objects from LD06 and IMC07. Symbols as in Fig.~\ref{fig:NOvsHe}. 
[\textit{See the electronic edition of the Journal for a color version of this figure.}]
\label{fig:ARvsO}}
\end {figure}

It is important to note the much larger dispersion in the distributions of He, O, N/O, and to a lesser extent He, in the LD06 and IMC07 data for objects in common, shown in Figs.~\ref{fig:NOvsHe}, \ref{fig:NOvsO}, and \ref{fig:ARvsO}. The reduction in the scatter in these diagrams is remarkable, and changes the interpretations that have been offered in the literature. Certainly our smaller range in O abundance is much more consistent with the average for \ion{H}{2} regions, and would not require some mechanism for O production during post-AGB nucleosynthesis, as LD06 suggested. The lower scatter in Fig.~\ref{fig:NOvsHe} leaves no doubt that there are no Type~I PNe in our sample, and if the scatter in the N/O vs. He/H diagrams of LD06 and \citet{MCI09} is dominated by large abundance errors, it would imply that the reported lack of segregation into Type~I and non-Type~I nebulae could largely be a result of large errors masking the underlying relationships. Similarly the evidence in Fig.~\ref{fig:NOvsO} that O is not destroyed in AGB stars that produce non-Type~I PNe would not have been convincing had the LD06 and IMC07 abundances been used for this study. Finally, our lower and more accurate S abundances compared to LD06 and \citet{MCI09} make a more convincing case for a genuine S anomaly.  

It is clear that careful attention to the calibration of deep optical spectra, along with near-IR spectra make it possible to achieve much higher levels of accuracy in the derived elemental abundances of PNe. Higher accuracy is very important for comparing abundances in PNe to the predicted chemical yields for the prior phase of AGB evolution. In a subsequent paper we will present similar optical and IR spectroscopy for a larger number of LMC PNe, and we will then be able to address nucleosynthetic yields from AGB stars more directly, and contrast them with these results for the SMC sample. 

\acknowledgements 

Partial support for this work was provided by NASA through grant GO-09077 from Space Telescope Science Institute, which is operated by the Association of Universities for Research in Astronomy, Incorporated, under NASA contract NAS5--26555. This work was also based in part on observations made with the \textit{Spitzer Space Telescope}, which is operated by the Jet Propulsion Laboratory, California Institute of Technology under a contract with NASA. Support was provided by NASA through an award issued by JPL/Caltech for program GO~20443. AGH and AM acknowledge support front grant AYA-2007-64748 from the Spanish Ministerio de Ciencia e Innovaci{\'o}n. The authors thank the anonymous referee for comments that helped to improve the content of this paper.

\clearpage


\clearpage


\begin{deluxetable}{llllcrlc}
\tabletypesize{\scriptsize}
\tablecolumns{7}
\tablewidth{0pt}
\tablecaption{Observing Log \label{tab:obsLog}}
\tablehead{
\colhead {} & \colhead {} & \colhead {T$_{exp}$} & \colhead {} &\colhead {Slit Width} & \colhead {Wavelength}  & \\
\colhead {Object} & \colhead {Date} & \colhead {(s)} & \colhead {Arm} &\colhead {(arcsec)} & \colhead {(\AA)} & \colhead {Blue Scale}
}
\startdata
SMC-MG~8    & 2003-Oct-28 & $2\times900$  & Blue & 1.5 & 3670--4130 & 0.70 \\*
            &             & $4\times900$  & Red  &     & 4000--7900 & \\
SMC-MG~13   & 2003-Oct-29 & $4\times900$  & Blue & 1.5 & 3670--4130 & 0.80 \\*
            &             & $4\times1200$ & Red  &     & 4000--7900 & \\
SMC-SMP~8   & 2003-Oct-30 & $2\times300$  & Blue & 1.5 & 3670--4130 & 0.65 \\*
            &             & $4\times600$  & Red  &     & 4000--7900 & \\
SMC-SMP~9   & 2003-Oct-30 & $4\times600$  & Blue & 1.5 & 3670--4130 & \nodata \\*
            &             & $2\times900$  & Red  &     & 4000--7900 & \\
SMC-SMP~11  & 2003-Oct-28 & $2\times1200$ & Blue & 1.5 & 3670--4130 & 0.60 \\*
            &             & 1200, 204 & Red  &     & 4000--7900 & \\
SMC-SMP~13  & 2003-Oct-28 & $2\times600$  & Blue & 1.5 & 3670--4130 & 0.97 \\*
            & 2003-Oct-29 & $2\times600$  & Red  & 1.5 & 4000--7900 & \\
SMC-SMP~14  & 2000-Dec-04 & $3\times900$  & Blue & 1.0 & 3480--3930 & 1.00 \\*
            &             & $4\times900$  & Red  &     & 3920--7630 & \\
SMC-SMP~17  & 2003-Oct-30 & $2\times300$  & Blue & 1.5 & 3670--4130 & 0.88 \\*
            &             & $2\times600$  & Red  &     & 4000--7900 & \\
SMC-SMP~18  & 2003-Oct-29 & $2\times480$  & Blue & 1.5 & 3670--4130 & 0.90 \\*
            &             & $2\times900$  & Red  &     & 4000--7900 & \\
SMC-SMP~19  & 2003-Oct-30 & $2\times480$  & Blue & 1.5 & 3670--4130 & 0.90 \\*
            &             & $2\times900$  & Red  &     & 4000--7900 & \\
SMC-SMP~20  & 2003-Oct-30 & $2\times300$  & Blue & 1.5 & 3670--4130 & 1.09 \\*
            &             & $2\times600$  & Red  &     & 4000--7900 & \\
SMC-SMP~23  & 2003-Oct-29 & $2\times420$  & Blue & 1.5 & 3670--4130 & 0.50 \\*
            &             & $4\times900$  & Red  &     & 4000--7900 & \\
SMC-SMP~24  & 2003-Oct-29 & $2\times420$  & Blue & 1.5 & 3670--4130 & 0.90 \\*
            &             & $2\times840$  & Red  &     & 4000--7900 & \\
SMC-SMP~27  & 2003-Oct-28 & $2\times600$  & Blue & 1.5 & 3670--4130 & 0.85 \\*
            &             & $2\times600$  & Red  &     & 4000--7900 & 
\enddata
\end{deluxetable}

\begin{deluxetable}{crrrrrrrrrrrrrrr}
\tabletypesize{\tiny}
\rotate
\tablecolumns{16}
\tablewidth{0pt}
\tablecaption{Optical Emission Line Intensities \label{tab:optFlux}}
\tablehead{
\colhead {} & \colhead {Wave} & \colhead {} & \colhead {} & \colhead {} & \colhead {} & \colhead {} & \colhead {} & \colhead {} & \colhead {} & \colhead {} & \colhead {} & \colhead {} & \colhead {} & \colhead {} & \colhead {} \\
\colhead {Ident.} & \colhead {(\AA)} & \colhead {MG~8} & \colhead {MG~13} & \colhead {SMP~8} & \colhead {SMP~9} & \colhead {SMP~11} & \colhead {SMP~13} & \colhead {SMP~14} & \colhead {SMP~17} & \colhead {SMP~18} & \colhead {SMP~19} & \colhead {SMP~20} & \colhead {SMP~23} & \colhead {SMP~24} & \colhead {SMP~27} 
}
\startdata
H16            & 3703.9 &    1.81 & \nodata &    1.89 & \nodata &    1.75 &    2.36 &    2.45 &    2.23 &    2.61 &    1.58 &    2.47 &    1.75 &    2.39 &    2.34 \\
H15            & 3712.0 &    1.84 & \nodata &    1.62 & \nodata &    1.73 &    1.85 &    2.23 &    1.78 &    1.94 &    1.22 &    2.05 &    1.68 &    1.89 &    1.72 \\
H14            & 3721.9 &    2.31 &    1.61 &    2.55 & \nodata &    2.77 &    2.47 &    3.17 &    2.33 &    2.76 &    2.22 &    2.58 &    2.08 &    2.81 &    2.31 \\
{[O~{\sc ii}]} & 3726.0 &  223.75 &    2.90 &    8.87 & \nodata &   63.86 &   14.37 &   33.40 &   15.48 &   27.52 &   16.01 &    3.61 &    3.21 &   39.72 &   10.67 \\
{[O~{\sc ii}]} & 3728.8 &  188.07 &    1.74 &    4.26 & \nodata &   37.47 &    6.75 &   17.37 &    7.21 &   11.80 &    8.70 &    1.53 &    2.08 &   18.61 &    4.59 \\
H13            & 3734.4 &    2.87 &    2.08 &    2.71 & \nodata &    1.94 &    2.54 &    3.52 &    2.41 &    2.58 &    2.46 &    3.09 &    2.52 &    2.56 &    2.43 \\
H12            & 3750.2 &    2.68 &    3.13 &    3.03 & \nodata &    1.66 &    3.36 &    3.80 &    2.85 &    3.46 &    2.82 &    3.48 &    3.18 &    3.06 &    2.86 \\
H11            & 3770.6 &    3.84 &    2.48 &    3.68 & \nodata &    2.39 &    3.84 &    5.02 &    3.72 &    3.98 &    3.49 &    3.82 &    3.72 &    3.70 &    3.58 \\
H10            & 3797.9 &    4.25 &    3.38 &    4.72 & \nodata &    3.75 &    4.95 &    6.57 &    4.61 &    4.92 &    4.58 &    4.70 &    4.40 &    4.84 &    4.67 \\
He~{\sc i}     & 3819.6 &    0.95 &    0.49 &    1.17 & \nodata &    0.39 &    1.22 &    0.00 &    1.10 &    1.15 &    0.74 &    1.28 &    1.24 &    1.12 &    1.17 \\

H9             & 3835.4 &    6.88 &    5.35 &    6.53 & \nodata &    4.49 &    6.25 &    9.16 &    6.11 &    6.63 &    5.94 &    6.56 &    6.19 &    6.46 &    6.22 \\
{[Ne~{\sc iii}]} & 3868.7 &  0.90 &   15.75 &   29.36 &   54.21 &    5.39 &   38.30 &   65.40 &   56.01 &   11.19 &   48.92 &   19.30 &   41.85 &   18.91 &   26.07 \\
H8             & 3889.1 &   14.83 &    8.58 &   16.43 &   20.11 &   10.41 &   13.35 &   18.92 &   16.44 &   13.02 &   15.36 &   13.45 &   18.89 &   15.45 &   15.83 \\
He~{\sc i}     & 3964.7 &    1.38 &    0.38 &    1.77 & \nodata &    0.49 &    0.75 & \nodata &    0.83 &    1.12 &    0.48 &    0.96 &    1.64 &    1.42 &    1.22 \\

{[Ne~{\sc iii}]} & 3967.4 & \nodata &  5.55 &    9.80 &   21.23 &    1.70 &   11.40 &   16.38 &   18.62 &    3.87 &   15.81 &    5.89 &   13.72 &    6.26 &    8.92 \\
H$\epsilon$    & 3970.1 &   13.40 &   12.03 &   14.21 &   13.65 &   10.67 &   12.62 &   12.88 &   13.14 &   14.28 &   12.93 &   14.92 &   14.15 &   15.20 &   12.90 \\
He~{\sc i}     & 4009.3 & \nodata &   15.04 &    0.35 & \nodata & \nodata &    0.21 & \nodata &    0.31 &    0.18 & \nodata & \nodata &    0.85 &    0.24 &    0.28 \\
He~{\sc i}     & 4026.4 &    1.32 &    0.89 &    2.23 & \nodata &    0.81 &    2.34 &    2.99 &    2.21 &    2.19 &    1.67 &    2.40 &    2.02 &    2.25 &    2.06 \\
{[S~{\sc ii}]} & 4068.6 & \nodata &    1.94 &    0.34 & \nodata &    0.86 &    1.14 &    2.04 &    1.23 &    1.04 &    1.73 &    0.44 &    0.96 &    0.85 &    0.51 \\

{[S~{\sc ii}]} & 4076.4 & \nodata & \nodata &    0.19 & \nodata &    0.33 &    0.36 &    0.53 &    0.31 &    0.48 &    0.00 &    0.19 &    0.00 &    0.45 &    0.18 \\
N~{\sc iii}    & 4097.3 & \nodata & \nodata & \nodata & \nodata & \nodata &    0.19 & \nodata &    0.31 & \nodata &    0.58 & \nodata & \nodata & \nodata &    0.21 \\
H$\delta$      & 4101.8 &   24.54 &   24.66 &   25.53 &   24.70 &   17.55 &   24.03 &   21.88 &   25.59 &   25.27 &   24.62 &   26.02 &   24.89 &   25.73 &   25.03 \\
C~{\sc ii}     & 4267.2 &    2.09 & \nodata &    1.10 & \nodata & \nodata & \nodata & \nodata &    1.08 &    0.80 &    0.63 &    0.89 & \nodata &    1.02 &    0.49 \\
H$\gamma$      & 4340.5 &   52.31 &   37.92 &   46.91 &   40.93 &   32.22 &   30.67 &   46.78 &   43.55 &   44.05 &   43.22 &   43.34 &   43.97 &   45.53 &   42.75 \\

{[O~{\sc iii}]} & 4363.2 &   1.88 &    6.35 &    8.63 &   15.00 &    3.18 &    9.07 &   12.18 &    9.59 &    3.21 &   13.29 &    6.52 &   15.15 &    4.36 &    7.49 \\
He~{\sc i}     & 4387.9 & \nodata & \nodata &    0.85 & \nodata & \nodata & \nodata & \nodata &    0.50 &    0.64 & \nodata &    0.68 & \nodata &    0.74 &    0.54 \\
He~{\sc i}     & 4471.5 &    5.40 & \nodata &    6.93 & \nodata & \nodata &    4.18 &    3.31 &    4.58 &    5.12 &    3.16 &    4.83 &    5.90 &    4.92 &    4.77 \\
C~{\sc iii}    & 4647.4 & \nodata & \nodata & \nodata & \nodata & \nodata & \nodata & \nodata &    0.30 & \nodata &    0.75 & \nodata &    0.64 & \nodata & \nodata \\
C~{\sc iii}    & 4650.2 & \nodata & \nodata & \nodata & \nodata & \nodata & \nodata & \nodata &    0.35 & \nodata &    0.56 & \nodata &    0.29 & \nodata & \nodata \\

He~{\sc ii}    & 4685.7 & 35.36\tablenotemark{a} &   96.36 & \nodata &   57.23 & \nodata & \nodata &   38.63 &    1.69 & \nodata &   43.82 & \nodata &    4.26 & \nodata & \nodata \\
{[Ar~{\sc iv}]} & 4711.3 & \nodata & \nodata &   0.44 & \nodata & \nodata & \nodata &    2.78 &    0.74 & \nodata &    1.32 & \nodata &    1.82 & \nodata &    0.34 \\
He~{\sc i}     & 4713.1 & \nodata &    2.65 &    1.00 &    2.16 & \nodata &    1.13 &    0.44 &    0.81 &    0.89 &    0.49 &    1.04 &    1.25 &    0.87 &    0.91 \\
{[Ne~{\sc iv}]} & 4725.6 & $<0.40$ & $<0.89$ & $<0.21$ & $<0.77$ & $<0.20$ & \nodata &   0.14 & $<0.08$ & $<0.10$ & $<0.27$ & $<0.07$ & $<0.28$ & $<0.10$ & $<0.07$ \\
{[Ar~{\sc iv}]} & 4740.2 & \nodata &   2.22 &    0.78 &   2.14 & \nodata &    1.28 &    2.64 &    0.97 & \nodata &    2.18 & \nodata &    2.26 & \nodata &    0.49 \\

He~{\sc ii}    & 4859.3 & \nodata &    4.66 & \nodata &    4.02 & \nodata & \nodata &    2.66 & \nodata & \nodata &    2.58 & \nodata & \nodata & \nodata & \nodata \\
H$\beta$       & 4861.3 &  100.00 &  100.00 &  100.00 &  100.00 &  100.00 &  100.00 &  100.00 &  100.00 &  100.00 &  100.00 &  100.00 &  100.00 &  100.00 &  100.00 \\
{[O~{\sc iii}]} & 4958.9 & \nodata & \nodata & \nodata & \nodata & \nodata & \nodata & 297.29 & \nodata & \nodata & \nodata & \nodata & \nodata & \nodata & \nodata \\
{[O~{\sc iii}]} & 5006.8 & 145.48 &  288.37 &  548.00 &  881.86 &  367.38 &  763.23 &  903.64 &  840.28 &  310.61 &  806.23 &  405.73 &  807.37 &  435.09 &  580.98 \\
He~{\sc i}     & 5015.0 &    2.24 & \nodata &    3.36 & \nodata & \nodata &    2.53 &    2.04 &    2.29 &    3.29 &    1.18 &    2.82 &    2.97 &    3.07 &    2.37 \\
He~{\sc i}     & 5047.7 & \nodata & \nodata & \nodata & \nodata & \nodata & \nodata & \nodata & \nodata &    0.30 & \nodata &    0.24 & \nodata &    0.36 &    0.23 \\

He~{\sc ii}    & 5411.5 & \nodata &   11.54 & \nodata &    5.35 & \nodata & \nodata &    3.77 &    0.17 & \nodata &    4.01 & \nodata &    0.53 & \nodata & \nodata \\
{[N~{\sc ii}]} & 5754.6 &    1.74 & \nodata &    0.26 &    0.43 & \nodata &    0.33 & \nodata &    0.26 &    1.04 &    0.23 &    0.31 &    0.00 &    0.51 &    0.17 \\
C~{\sc iv}     & 5801.5 & \nodata &    1.88 & \nodata & \nodata & \nodata & \nodata & \nodata & \nodata & \nodata &    0.62 & \nodata & \nodata & \nodata & \nodata \\
C~{\sc iv}     & 5812.0 & \nodata &    1.24 & \nodata & \nodata & \nodata & \nodata & \nodata & \nodata & \nodata &    0.39 & \nodata & \nodata & \nodata & \nodata \\
He~{\sc i}     & 5875.7 &   14.36 &    1.56 &   15.72 &    7.05 &   19.23 &   16.51 &   12.75 &   16.40 &   14.05 &   10.92 &   17.03 &   16.13 &   15.90 &   16.78 \\

{[O~{\sc i}]}  & 6300.3 &    8.17 & \nodata &    3.75 &    1.53 &    3.77 &    3.17 &    3.30 &    2.33 &    1.41 &    3.01 &    1.34 & \nodata &    1.69 &    1.12 \\
{[S~{\sc iii}]} & 6312.1 &   0.93 & \nodata &    0.72 &    2.51 &    3.79 &    0.78 &    0.92 &    0.62 &    0.84 &    0.67 &    0.37 & \nodata &    0.76 &    0.59 \\
{[O~{\sc i}]}  & 6363.8 &    2.45 & \nodata &    1.26 &    0.86 & \nodata &    1.06 &    1.20 &    0.81 &    0.54 &    0.96 &    0.46 & \nodata &    0.54 &    0.40 \\
{[Ar~{\sc v}]} & 6435.0 & \nodata & \nodata & \nodata & \nodata & \nodata & \nodata &    0.45 & \nodata & \nodata & \nodata & \nodata & \nodata & \nodata & \nodata \\
C~{\sc ii}     & 6462.0 & \nodata & \nodata & \nodata & \nodata & \nodata & \nodata & \nodata &    0.13 &    0.12 & \nodata &    0.12 & \nodata &    0.10 &    0.08 \\
{[N~{\sc ii}]} & 6548.0 &   33.08 & \nodata &    1.47 &    8.17 &    8.35 &    3.21 &    4.35 &    2.74 &    7.35 &    3.50 &    1.00 &    0.77 &    6.84 &    1.88 \\

He~{\sc ii}    & 6560.1 &    0.63 &   25.02 & \nodata &   10.94 & \nodata & \nodata &    8.63 & \nodata & \nodata &    7.50 & \nodata &    1.61 & \nodata & \nodata \\
H$\alpha$      & 6562.8 &  314.14 &  294.23 &  284.21 &  307.06 &  612.90 &  300.46 &  305.50 &  303.31 &  293.03 &  322.68 &  300.93 &  303.18 &  295.45 &  311.88 \\
C~{\sc ii}     & 6578.0 &    1.66 & \nodata &    0.79 & \nodata & \nodata &    0.40 & \nodata &    0.65 &    0.85 &    0.30 &    0.80 &    0.37 &    0.65 &    0.43 \\
{[N~{\sc ii}]} & 6583.4 &   92.45 &    2.02 &    3.74 &   24.29 &   25.97 &    9.34 &   13.37 &    8.29 &   21.66 &    8.49 &    3.26 &    1.35 &   18.97 &    5.71 \\
He~{\sc i}     & 6678.2 &    5.16 & \nodata &    4.54 &    2.16 &    7.21 &    4.61 &    3.66 &    4.66 &    4.03 &    3.20 &    4.77 &    5.25 &    4.27 &    4.79 \\

{[S~{\sc ii}]} & 6716.5 &   12.61 & \nodata &    0.42 &    8.70 &    9.12 &    0.83 &    2.02 &    0.84 &    0.56 &    1.28 &    0.00 &    0.00 &    1.22 &    0.48 \\
{[S~{\sc ii}]} & 6730.9 &   14.46 & \nodata &    0.80 &    8.06 &   11.55 &    1.46 &    3.07 &    1.48 &    1.03 &    1.85 &    0.11 &    0.40 &    2.18 &    0.89 \\
{[Ar~{\sc v}]} & 7005.7 & \nodata &    2.21 & \nodata & \nodata & \nodata & \nodata &    0.53 & \nodata & \nodata &    0.58 & \nodata & \nodata & \nodata & \nodata \\
He~{\sc i}     & 7065.3 &    4.29 & \nodata &    8.81 &    2.40 &   19.15 &   11.65 &    4.61 &    9.84 &   10.03 &    5.11 &   13.12 &    6.62 &    9.14 &   11.29 \\
{[Ar~{\sc iii}]} & 7135.9 &  6.83 &    3.79 &    4.48 &   10.56 &   22.18 &    5.35 &    6.40 &    4.58 &    6.25 &    4.20 &    2.20 &    3.27 &    4.91 &    3.53 \\

C~{\sc ii}     & 7236.2 &    1.08 &    1.79 &    0.55 & \nodata & \nodata &    0.49 & \nodata &    0.70 &    0.84 &    0.35 &    0.95 & \nodata &    0.70 &    0.47 \\
O~{\sc i}      & 7254.4 & \nodata & \nodata & \nodata & \nodata & \nodata & \nodata & \nodata & \nodata &    0.31 & \nodata &    0.37 & \nodata & \nodata & \nodata \\
He~{\sc i}     & 7281.0 &    1.44 & \nodata &    1.67 &    1.03 &    2.82 &    1.24 & \nodata &    1.10 &    1.12 &    0.80 &    1.32 &    1.47 &    1.32 &    1.27 \\
{[O~{\sc ii}]} & 7319.4 &   15.62 & \nodata &    1.91 &    2.93 &   30.72 &    4.28 &    2.71 &    2.93 &   12.41 &    3.00 &    4.92 &    0.86 &    6.53 &    2.67 \\
{[O~{\sc ii}]} & 7329.9 &   13.57 & \nodata &    1.57 &    2.73 &   24.69 &    3.48 &    2.46 &    2.28 &    9.67 &    2.58 &    3.79 &    0.67 &    5.17 &    2.12 

\enddata
\tablenotetext{a}{Emission is entirely stellar in origin.} 
\tablecomments{{Line} intensities scaled to $I$(H$\beta$) = 100, uncorrected for reddening.}
\end{deluxetable}

\begin{deluxetable}{crrrrrrrrrrrrrr}
\tabletypesize{\scriptsize}
\rotate
\tablecolumns{14}
\tablewidth{0pt}
\tablecaption{Infrared Emission Line Intensities \label{tab:irFlux}}
\tablehead{
\colhead {} & \colhead {Wave} & \colhead {} & \colhead {} & \colhead {} & \colhead {} & \colhead {} & \colhead {} & \colhead {} & \colhead {} & \colhead {} & \colhead {} & \colhead {} & \colhead {} \\
\colhead {Ident.} & \colhead {($\mu$m)} & \colhead {$f_\lambda$} & \colhead {SMP~8} & \colhead {SMP~11\tablenotemark{a}} & \colhead {SMP~13} & \colhead {SMP~14} & \colhead {SMP~17} & \colhead {SMP~18} & \colhead {SMP~19} & \colhead {SMP~20} & \colhead {SMP~23} & \colhead {SMP~24\tablenotemark{a}} & \colhead {SMP~27} 
}
\startdata
H$\beta$        & 0.4861 &   0.0000  & $-12.81$ & $-13.13$ & $-12.59$ & $-13.04$ & $-12.55$ & $-12.66$ & $-13.04$ & $-12.47$ & $-13.18$ & $-12.66$ & $-12.51$ \\
{[Ar~{\sc ii}]}  &  6.98 & $-0.9779$ &  $<32.8$ &  \nodata &  $<11.5$ &  $<44.5$ &  $<12.9$ &  $<19.2$ &  $<47.2$ &  $<23.8$ &  $<78.9$ &  \nodata &  $<11.1$ \\
{[Ne~{\sc vi}]}  &  7.65 & $-0.9799$ &  $<18.1$ &  \nodata &  $<17.5$ &  $<78.6$ &  $<39.7$ &  $<23.1$ &  $<81.0$ &  $<17.5$ &  $<39.5$ &  \nodata &  $<11.0$ \\
{[Ar~{\sc iii}]} &  8.99 & $-0.9829$ &   $<5.9$ &  \nodata &   $<5.9$ &  $<21.0$ &   $<8.1$ &  $<13.6$ &  $<19.1$ &   $<4.2$ &  $<12.4$ &  \nodata &   $<3.5$ \\
{[S~{\sc iv}]}   & 10.53 & $-0.9854$ &     18.3 &     50.7 &     20.2 &     47.3 &     29.8 &   $<4.9$ &     39.8 &   $<5.5$ &     42.0 &      7.0 &     13.6 \\
{[Ne~{\sc ii}]}  & 12.82 & $-0.9881$ &   $<3.6$ &    181.3 &   $<3.5$ &  $<10.1$ &   $<6.8$ &     20.9 &   $<7.4$ &   $<3.0$ &  $<12.3$ &      8.3 &   $<2.8$ \\
{[Ne~{\sc v}]}   & 14.29 & $-0.9893$ &   $<7.6$ &   $<5.0$ &   $<3.5$ &     19.1 &   $<3.3$ &   $<6.3$ &     28.8 &   $<4.7$ &  $<18.6$ &   $<1.1$ &   $<3.2$ \\
{[Ne~{\sc iii}]} & 15.56 & $-0.9900$ &     43.2 &    110.5 &     56.7 &     74.3 &     89.3 &     24.1 &     82.3 &     16.2 &     52.3 &     34.1 &     33.8 \\
{[S~{\sc iii}]}  & 18.68 & $-0.9918$ &      9.0 &    170.9 &   $<7.1$ &  $<10.3$ &      7.3 &     12.0 &  $<15.4$ &  $<11.9$ &  $<20.9$ &     10.4 &      4.4 \\
{[Ne~{\sc v}]}   & 24.23 & $-0.9937$ &   $<3.2$ &  \nodata &  $<13.0$ &     23.5 &   $<5.4$ &   $<9.0$ &     32.2 &   $<2.8$ &  $<13.3$ &  \nodata &   $<3.8$ \\
{[O~{\sc iv}]}   & 25.93 & $-0.9941$ &   $<3.2$ &  $<15.9$ &   $<2.9$ &    264.4 &   $<3.4$ &   $<5.2$ &    277.1 &   $<2.6$ &     11.0 &   $<3.8$ &   $<3.7$ \\
{[S~{\sc iii}]}  & 33.64 & $-0.9954$ &  $<13.2$ &  \nodata &   $<7.1$ &  $<14.1$ &      5.3 &  $<17.3$ &  $<19.4$ &   $<4.9$ &  $<40.1$ &  \nodata &   $<7.7$ \\
{[Si~{\sc ii}]}  & 34.84 & $-0.9956$ &  $<15.7$ &  \nodata &  $<10.9$ &  $<18.1$ &   $<8.5$ &  $<13.0$ &  $<20.7$ &   $<6.8$ &  $<30.4$ &  \nodata &   $<9.1$ \\
{[Ne~{\sc iii}]} & 36.01 & $-0.9958$ &  $<28.9$ &  \nodata &   $<9.0$ &  $<93.5$ &  $<29.1$ &  $<21.6$ &  $<70.7$ &   $<9.0$ & $<106.6$ &  \nodata &  $<41.1$ \\
\enddata
\tablecomments{{Line} intensities are scaled to $I$(H$\beta$) = 100, uncorrected for reddening; log $F$(H$\beta$) in erg~s$^{-1}$~cm$^{-2}$, given in the first row, are from Stanghellini et al. (2003).}
\tablenotetext{a}{Adapted from \citet{BernardSalas08}.}
\end{deluxetable}

\begin{deluxetable}{lrrrrrllrr}
\tabletypesize{\scriptsize}
\rotate
\tablecolumns{10}
\tablewidth{0pt}
\tablecaption{Physical Diagnostics \label{tab:Diag}}
\tablehead{
\colhead {} & \colhead {} & \colhead {} & \colhead {} & \colhead {} & \colhead{} & \colhead {} & \colhead {} & \multicolumn{2}{c}{Adopted T$_e$}  \\
\cline{9-10} \\
\colhead {} & \colhead {} & \colhead {N$_e$[O~{\sc ii}]} & \colhead {N$_e$[S~{\sc ii}]} & \colhead {N$_e$[Ar~{\sc iv}]} & \colhead {Adopted N$_e$}  & \colhead {T$_e$[N~{\sc ii}]} & \colhead {T$_e$[O~{\sc iii}]} & \colhead {Low} & \colhead {Medium} \\
\colhead {Object} & \colhead {$c$\tablenotemark{a}} & \colhead {(cm$^{-3}$)} & \colhead {(cm$^{-3}$)} &
\colhead {(cm$^{-3}$)} & \colhead {(cm$^{-3}$)} & \colhead {(K)} & \colhead {(K)} & \colhead {(K)} & \colhead {(K)} 
}
\startdata
MG~8   & 0.144 &     670 &    1130 & \nodata & 1100 & $11520^{+1620}_{-1240}$ & $13000^{+1100}_{-960}$  &   11500 & 13000  \\
MG~13  & 0.060\tablenotemark{b} &    1690 & \nodata & \nodata & 1700 &                 \nodata & $16200^{+1740}_{-1440}$ & \nodata & 16200  \\
SMP~8  & 0.026 &    2770 &   9050: & 19980: & 2800 &                 \nodata & $13700^{+1170}_{-1110}$ & \nodata & 13700  \\
SMP~9  & 0.070 & \nodata &     470 & \nodata &  470 & $11300^{+1580}_{-1140}$ & $14400^{+1440}_{-1110}$ &   11300 & 14400  \\
SMP~11 & 1.020\tablenotemark{b} &  1650 &  1590 & \nodata & 1600 &                 \nodata & $12550^{+1070}_{-870}$  & \nodata & 12550  \\
SMP~13 & 0.190 &    2900 &  12700: & \nodata & 2900 &                 \nodata & $12800^{+1080}_{-940}$  & \nodata & 12800  \\
SMP~14 & 0.069 &    2230 &    2750 &    3890 & 2400 &                 \nodata & $13100^{+1090}_{-990}$  & \nodata & 13100  \\
SMP~17 & 0.064 &    2880 &    5720 &  10000: & 2900 &                 \nodata & $12200^{+980}_{-860}$   & \nodata & 12200  \\
SMP~18 & 0.122 &    3590 &    7550 & \nodata & 3600 &                 \nodata & $11860^{+800}_{-870}$   & \nodata & 11800  \\
SMP~19 & 0.161 &    2070 &    2680 &  17650: & 2100 & $13125^{+2110}_{-1620}$ & $14380^{+1330}_{-1190}$ &   13100 & 14400  \\
SMP~20 & 0.000 &    3880 & \nodata & \nodata & 3900 &                 \nodata & $13820^{+1380}_{-1010}$ & \nodata & 13800  \\
SMP~23 & 0.101 &    1390 & \nodata &   9500: & 1400 &                 \nodata & $15100^{+1380}_{-1350}$ & \nodata & 15100  \\
SMP~24 & 0.047 &    2780 &    6100 & \nodata & 2800 & $13140^{+2040}_{-1680}$ & $11620^{+910}_{-740}$   & \nodata & 11600  \\
SMP~27 & 0.040 &    3650 &    7910 &  12860: & 3650 & $13890^{+2460}_{-1740}$ & $12730^{+1180}_{-820}$  & \nodata & 12700  \\
\enddata
\tablenotetext{a}{Extinction constants from \citet{Stang_etal03} except where noted.}
\tablenotetext{b}{Extinction derived from this work.}
\end{deluxetable}

\begin{deluxetable}{lrrrrrrrrrrrrrr}
\tabletypesize{\tiny}
\rotate
\tablecolumns{15}
\tablewidth{0pt}
\tablecaption{Ionic Abundances \label{tab:Ionic}}
\tablehead{
\colhead {} & \colhead {MG~8} & \colhead {MG~13} & \colhead {SMP~8} & \colhead {SMP~9} & \colhead {SMP~11} & \colhead {SMP~13} & \colhead {SMP~14} & \colhead {SMP~17} & \colhead {SMP~18} & \colhead {SMP~19} & \colhead {SMP~20} & \colhead {SMP~23} & \colhead {SMP~24} & \colhead {SMP~27} 
}
\startdata
He$^{+}$     &    0.120 &   0.013  &    0.130 &    0.064 &    0.102 &    0.128 &    0.102 &    0.137 &    0.114 &    0.085 &    0.139 &    0.132 &    0.136 &    0.139 \\
He$^{+2}$    &    0.000 &   0.085  &    0.000 &    0.049 &    0.000 &    0.000 &    0.033 &    0.001 &    0.000 &    0.038 &    0.000 &    0.004 &    0.000 &    0.000 \\
\tableline
N$^{+}$      & 1.19(-5) & 1.32(-7) & 3.79(-7) & 3.34(-6) & 1.47(-6) & 1.04(-6) & 1.33(-6) & 9.86(-7) & 2.71(-6) & 8.41(-7) & 3.21(-7) & 1.14(-7) & 2.62(-6) & 6.33(-7) \\
ICF(N)       & 1.19     & 236.     & 29.9     & 5.29     & 2.27     & 19.0     & 17.3     & 24.3     & 4.59     & 22.9     & 27.8     & 120.     & 5.61     & 23.5 \\
\tableline
O$^{+}$      & 1.19(-4) & 4.26(-7) &   2.55(-6) & 3.97(-5) & 4.60(-5)   & 6.01(-6)   & 1.01(-5) & 6.75(-6) & 1.71(-5)   & 5.57(-6) & 1.99(-6)   & 7.16(-7) & 2.04(-5)   & 4.27(-6) \\
O$^{+2}$     & 2.24(-5) & 2.52(-5) &   7.35(-5) & 1.04(-4) & 5.85(-5)   & 1.09(-4)   & 1.35(-4) & 1.56(-4) & 6.18(-5)   & 9.38(-5) & 5.35(-5)   & 8.36(-5) & 9.40(-5)   & 9.61(-5) \\
O$^{+3}$ IR  & \nodata  & \nodata  & $<7.1$(-7) & \nodata  & $<3.2$(-7) & $<4.9$(-7) & 5.02(-5) & 7.43(-7) & $<1.4$(-6) & 3.99(-5) & $<6.2$(-7) & 1.64(-6) & $<9.4$(-7) & $<8.0$(-7) \\
ICF(O)       & 1.00     & 3.91     &   1.00     & 1.47     & 1.00       & 1.00       & 1.00     & 1.00     & 1.00       & 1.00     & 1.00       & 1.00     & 1.00       & 1.00 \\
\tableline
Ne$^{+}$ IR  & \nodata  & \nodata  & $<3.8$(-6) & \nodata  & 2.11(-5)   & $<2.7$(-6) & $<9.5$(-6) & $<6.9$(-6) & 2.34(-5)   & $<5.9$(-6) & $<2.9$(-6) & $<1.1$(-5) & 9.29(-6)   & $<2.5$(-6) \\
Ne$^{+2}$    & 3.85(-7) & 3.28(-6) & 1.06(-5)   & 1.60(-5) &   4.44(-6) & 1.69(-5)   & 2.25(-5)   & 3.01(-5)   & 7.00(-6)   & 1.60(-5)   & 6.26(-6)   & 1.16(-5)   & 1.20(-5)   & 1.23(-5) \\
Ne$^{+2}$ IR & \nodata  & \nodata  & 2.28(-5)   & \nodata  & 6.24(-6)   & 2.13(-5)   & 3.45(-5)   & 4.42(-5)   & 1.34(-5)   & 3.08(-5)   & 7.78(-6)   & 2.28(-5)   & 1.87(-5)   & 1.51(-5) \\
Ne$^{+4}$ IR & \nodata  & \nodata  & $<2.9$(-7) & \nodata  & $<3.5$(-8) & $<1.7$(-7) & 1.12(-6)   & $<2.1$(-7) & $<4.4$(-7) & 1.42(-6)   & $<2.7$(-7) & $<8.5$(-7) & $<7.6$(-8) & $<1.9$(-7) \\
Ne$^{+5}$ IR & \nodata  & \nodata  & $<1.5$(-6) & \nodata  & \nodata    & $<1.0$(-6) & $<5.8$(-6) & $<3.0$(-6) & $<1.9$(-6) & $<5.1$(-6) & $<1.3$(-6) & $<3.0$(-6) & \nodata    & $<7.6$(-7) \\
ICF(Ne)      & 6.32     & 3.98     & 1.03       & 2.03     &   1.79     & 1.06       & 1.30       & 1.05       & 1.28       & 1.36       & 1.04       & 1.03       & 1.22       & 1.04 \\
\tableline
S$^{+}$      & 4.89(-7) & \nodata & 2.37(-8) & 2.98(-7) & 1.78(-7) & 5.81(-8)   & 1.10(-7)   & 6.70(-8) & 5.77(-8)   & 6.77(-8)   & 9.04(-9)   & 1.54(-8)   & 9.14(-8) & 3.39(-8) \\
S$^{+2}$     & 7.46(-7) & \nodata  & 5.12(-7) & 1.52(-6) & 2.03(-6) & 4.86(-7)   & 7.48(-7)   & 6.48(-7) & 9.22(-7)   & 3.79(-7)   & 2.63(-7)   & \nodata    & 9.69(-7) & 5.35(-7) \\
S$^{+2}$  IR & \nodata  & \nodata  & 8.02(-7) & \nodata  & 1.58(-6) & $<4.6$(-7) & $<8.0$(-7) & 6.32(-7) & 1.23(-6)   & $<9.4$(-7) & $<1.0$(-6) & $<1.4$(-6) & 1.01(-6) & 3.59(-7) \\
S$^{+3}$  IR & \nodata  & \nodata  & 4.52(-7) & \nodata  & 1.31(-7) & 3.59(-7)   & 1.02(-6)   & 6.98(-7) & $<1.3$(-7) & 6.86(-7)   & $<1.3$(-7) & 8.26(-7)   & 1.80(-7) & 2.92(-7) \\
ICF(S)       & 1.00     & \nodata  & 1.00     & 1.21     & 1.00     & 1.00       & $>1.00$    & 1.00     & 1.00       & $>1.00$    & 1.70       & $>3.4$     & 1.00     & 1.00 \\
\tableline
Ar$^{+}$  IR & \nodata  & \nodata  & $<2.2$(-6) & \nodata  & \nodata  & $<5.5$(-7) & $<2.6$(-6) & $<8.1$(-7) & $<1.4$(-6) & $<2.4$(-6) & $<1.4$(-6) & $<4.3$(-6) & \nodata  & $<6.2$(-7) \\
Ar$^{+2}$    & 3.17(-7) & 1.28(-7) & 2.08(-7)   & 4.32(-7) & 5.27(-7) & 2.85(-7)   & 3.12(-7)   & 2.58(-7)   & 3.53(-7)   & 1.59(-7)   & 1.03(-7)   & 1.20(-7)   & 3.12(-7) & 1.87(-7) \\
Ar$^{+2}$ IR & \nodata  & \nodata  & $<4.8$(-7) & \nodata  & \nodata  & $<3.4$(-7) & $<1.5$(-6) & $<6.2$(-7) & $<1.2$(-6) & $<1.1$(-6) & $<3.1$(-7) & $<8.2$(-7) & \nodata  & $<2.4$(-7) \\
Ar$^{+3}$    & \nodata  & 1.41(-7) & 5.02(-8)   & 1.80(-7) & \nodata  & \nodata    & 2.44(-7)   & 9.23(-8)   & \nodata    & 1.31(-7)   & \nodata    & 1.37(-7)   & \nodata  & 4.04(-8) \\
Ar$^{+4}$    & \nodata  & 1.35(-7) & \nodata    & \nodata  & \nodata  & \nodata    & 6.79(-8)   & \nodata    & \nodata    & 4.17(-8)   & \nodata    & \nodata    & \nodata  & \nodata  \\
ICF(Ar)      & 6.32     & 1.00     & 1.03       & 1.23     & 1.79     & 1.00       & 1.06       & 1.04       & 1.28       & 1.05       & 1.04       & 1.01       & 1.22     & 1.04 \\

\enddata
\tablecomments{ {a}($-b$) $\equiv$ a $\times 10^{-b}$}
\end{deluxetable}

\begin{deluxetable}{lrrrrrrr}
\tabletypesize{\footnotesize}
\tablecolumns{8}
\tablewidth{0pt}
\tablecaption{Elemental Abundances \label{tab:Abund}}
\tablehead{
\colhead {Object} & \colhead {log He} & \colhead {log N} & \colhead {log O} & \colhead {log Ne} & 
\colhead {log S} & \colhead {log Ar} & \colhead {log N/O} 
}
\startdata
MG~8   & $11.08\pm0.04$ & $7.15\pm0.13$ & $8.15 \pm0.19$& $6.39\pm0.18$ & $6.09\pm0.14$ & $6.30\pm0.10$ & $-1.00$ \\
MG~13  & $10.99\pm0.04$ & $>5.12$ & $8.00\pm0.11$ & $7.12\pm0.12$ & \nodata & $5.61\pm0.11$ & $-0.51$\tablenotemark{a} \\
SMP~8  & $11.11\pm0.04$ & $7.05\pm0.11$ & $7.88\pm0.11$ & $7.37\pm0.05$ & $6.11\pm0.05$ & $5.43\pm0.12$ & $-0.83$  \\
SMP~9  & $11.05\pm0.04$ & $7.25\pm0.15$ & $8.32\pm0.16$ & $7.51\pm0.12$ & $6.34\pm0.14$ & $5.88 \pm0.12$& $-1.08$  \\
SMP~11 & $11.01\pm0.04$ & $6.52\pm0.09$ & $8.02\pm0.12$ & $6.90\pm0.05$ & $6.28\pm0.05$ & $5.97\pm0.08$ & $-1.50$  \\
SMP~13 & $11.11\pm0.04$ & $7.30\pm0.11$ & $8.06\pm0.11$ & $7.35\pm0.05$ & $5.96\pm0.16$ & $5.46\pm0.11$ & $-0.76$  \\
SMP~14 & $11.13\pm0.04$ & $7.36\pm0.09$ & $8.29\pm0.09$ & $7.65\pm0.05$ & $>6.27$ & $5.82\pm0.11$ & $-0.93$  \\
SMP~17 & $11.14\pm0.04$ & $7.38\pm0.09$ & $8.21\pm0.11$ & $7.67\pm0.05$ & $>6.15$ & $5.56 \pm0.11$& $-0.83$  \\
SMP~18 & $11.06\pm0.04$ & $7.11\pm0.10$ & $7.90\pm0.12$ & $7.57\pm0.05$ & $6.18\pm0.05$ & $5.67\pm0.10$ & $-0.79$  \\
SMP~19 & $11.09\pm0.04$ & $7.28\pm0.16$ & $8.14\pm0.10$ & $7.62\pm0.05$ & $>6.05$ & $5.54\pm0.11$ & $-0.86$  \\
SMP~20 & $11.14\pm0.04$ & $6.95\pm0.12$ & $7.74\pm0.11$ & $6.91\pm0.05$ & $5.67\pm0.18$ & $5.03\pm0.11$ & $-0.79$  \\
SMP~23 & $11.13\pm0.04$ & $7.14\pm0.16$ & $7.93\pm0.11$ & $7.37\pm0.05$ & $>5.93$ & $5.41\pm0.11$ & $-0.80$  \\
SMP~24 & $11.13\pm0.04$ & $7.17\pm0.09$ & $8.06\pm0.11$ & $7.36\pm0.05$ & $6.11\pm0.05$ & $5.58\pm0.11$ & $-0.89$  \\
SMP~27 & $11.14\pm0.04$ & $7.17\pm0.12$ & $8.00\pm0.11$ & $7.20\pm0.05$ & $5.84\pm0.05$ & $5.38 \pm0.11$& $-0.83$  \\ \hline \\
Median & 11.11 & 7.17 & 8.04 & 7.40 & 6.11 & 5.57 & $-0.83$ \\
Median (LD06) & 10.95 & 7.03 & 8.11 & 7.18 & 6.77 & 5.65 & $-1.05$ \\
Median (IMC07) & 10.95 & 7.26 & 8.10 & 7.35 & 6.72 & 5.63 & $-0.90$ \\
Mean H~{\sc ii}\tablenotemark{b} & 10.90 & 6.46 & 7.97 & 7.22 & 6.32 & 5.78 & $-1.51$ 
\enddata
\tablecomments{{Abundances} of element $X$ expressed as $12 + $log N($X$)/N(H$^+$)}
\tablenotetext{a}{Based on N$^+$/O$^+$.}
\tablenotetext{b}{Data from the compilation of \citet{Den89}.}
\end{deluxetable}


\end{document}